\documentclass[times,twocolumn,final]{elsarticle}
\usepackage{prletters}

\usepackage{hyperref}
\usepackage{setspace}

\journal{Pattern Recognition Letters}

\usepackage[utf8]{inputenc}
\usepackage[linesnumbered]{algorithm2e}
\usepackage{amsmath}
\usepackage{amsfonts}
\usepackage{booktabs}
\usepackage{dirtytalk}
\usepackage{enumerate}
\usepackage{cleveref}
\usepackage{paralist}
\usepackage{multicol}
\usepackage{multirow}
\usepackage{pifont}
\usepackage[figuresright]{rotating}
\usepackage{threeparttable}
\usepackage{tablefootnote}
\usepackage{wrapfig}

\Crefname{algocf}{Algorithm}{Algorithms}

\usepackage[colorinlistoftodos,prependcaption,textsize=tiny]{todonotes}

\usepackage[%
  acronym, shortcuts,
  xindy={language=german-duden,codepage=utf8}
]{glossaries}

\renewcommand*{\CustomAcronymFields}{%
  name={\the\glsshorttok},
  description={\the\glslongtok},
}

\SetCustomStyle

\newcommand{\etal}{et al.}
\newcommand{\ie}{i.e.,}
\newcommand{\eg}{e.g.,}

\newcommand{\hea}{\textit{.hea}}
\newcommand{\m}{\textit{.m}}
\newcommand{\cpp}{\textit{.cpp}}
\newcommand{\atr}{\textit{.atr}}
\newcommand{\dat}{\textit{.dat}}

\newacronym{BAN}{BAN}{Body Area Network}
\newacronym{BCH}{BCH}{Bose-Chaudhuri-Hocquenghem}
\newacronym{BP}{BP}{Blood Pressure}
\newacronym{bpm}{bpm}{beats per minute}
\newacronym{BSN}{BSN}{Body Sensor Networks}
\newacronym{CWT}{CWT}{Continuous Wavelet Transform}
\newacronym{DCTAC}{DCT/AC}{Discrete Cosine Transform of the windowed Auto-Correlations}
\newacronym{DFT}{DFT}{Discrete Fourier Transform}
\newacronym{DWT}{DWT}{Discrete Wavelet Transform}
\newacronym{ECC}{ECC}{Error Correcting Code}
\newacronym{ECG}{ECG}{Electrocardiogram}
\newacronym{EEG}{EEG}{Electroencephalogram}
\newacronym{EMG}{EMG}{Electromyogram}
\newacronym{EOG}{EOG}{Electrooculography}
\newacronym{EI}{EI}{Entity Recognition}
\newacronym{FIR}{FIR}{Finite Impulse Response}
\newacronym{FSM}{FSM}{Finite State Machine}
\newacronym{FT}{FT}{Fourier Transform}
\newacronym{FFT}{FFT}{Fast Fourier Transform}
\newacronym{GSR}{GSR}{Galvanic Skin Response}
\newacronym{GPS}{GPS}{Global Positioning System}
\newacronym{HBBS}{HBBS}{Heart-Beat-Based Security}
\newacronym{HRV}{HRV}{Heart-Rate Variability}
\newacronym{ICU}{ICU}{Intensive Care Unit}
\newacronym{ICD}{ICD}{Implantable Cardiac Defibrillators}
\newacronym{IIR}{IIR}{Infinite Impulse Response}
\newacronym{IMD}{IMD}{Implantable Medical Device}
\newacronym{IOT}{IoT}{Internet of Things}
\newacronym{IPI}{IPI}{Inter-Pulse Interval}
\newacronym{LFSR}{LFSR}{Linear Feedback Shift Register}
\newacronym{LED}{LED}{Light Emitting Diode}
\newacronym{LSB}{LSB}{Least Significant Bit}
\newacronym{LPC}{LPC}{Linear Prediction Coding}
\newacronym{LRS}{LRS}{Longest Repeated Substring}
\newacronym{MCW}{MCW}{Most Common in Window}
\newacronym{MMC}{MMC}{Markov Model with Counting}
\newacronym{NIST}{NIST}{National Institute of Standard and Technology}
\newacronym{NISTSTS}{NIST STS}{National Institute of Standard and Technology Statistical Test Suite}
\newacronym{PAN}{PAN}{Personal Area Network}
\newacronym{PCG}{PCG}{Phonocardiogram}
\newacronym{PPG}{PPG}{Photoplethysmographic}
\newacronym{PRNG}{PRNG}{Pseudorandom Number Generators}
\newacronym{PUF}{PUF}{Physical Unconlable Function}
\newacronym{RF}{RF}{Radio Frequency}
\newacronym{RMSE}{RMSE}{Root-Mean-Square Error}
\newacronym{RVE}{RVE}{Runtime Verification Extractor}
\newacronym{STFT}{STFT}{Short Time Fourier Transform}
\newacronym{TCTL}{TCTL}{Timed Computation Tree Logic}
\newacronym{TSCH}{TSCH}{Time Slotted Channel Hopping}
\newacronym{VANET}{VANET}{Vehicular Ad-Hoc Network}
\newacronym{VAR}{VAR$_{is}$}{Inter-Sensor Variability}
\newacronym{WBAN}{WBAN}{Wireless Body Area Network}
\newacronym{WT}{WT}{Wavelet Transform} 

\AtBeginDocument{%
  \providecommand\BibTeX{{%
    \normalfont B\kern-0.5em{\scshape i\kern-0.25em b}\kern-0.8em\TeX}}}

\makeatletter
\begingroup \lccode`+=32 \lowercase
 {\endgroup \def\Url@ObeySp{\Url@Edit\Url@String{ }{+}}}
 \def\Url@space{\penalty\Url@sppen\ }
\makeatother

\begin{document}
\begin{frontmatter}

\title{Heart-Based Biometric Protocols: A look back over almost two decades}

\author{ Lara Ortiz-Martin and
       Pablo Picazo-Sanchez 
}





\begin{abstract}
This article surveys the literature over the period 2003-2021 on heart-based biometric protocols. In particular, we focus on how the heart signal is transformed from a continuous wave to discrete values to be used afterwards in authentication protocols. 
We explain and classify the surveyed proposals according to three main parameters:  
\begin{inparaenum}[i)]
    \item the dataset they use for testing their results;   
    \item the delineation algorithms they use to extract the fiducial points, and;
    \item the cryptographic tests they run (if any) to validate how random the extracted token is.
\end{inparaenum} 
\end{abstract}



\end{frontmatter}


\section{Introduction}
Over 70\% of the population in the US have used some kind of biometric system, ranging from unlocking the smartphone, generating a \gls*{ECG} using a smartwatch, or in the boarding control of the airports \cite{statsBiometric}. In this paper, we focus on how the heart signal as a biological source can be used in information security, not only to create biometric systems but also as part of cryptographic protocols. In more detail, we classify and review the main proposals existing in the literature between 2003 and 2021. 

Traditionally, to get an \gls*{ECG}, a patient had to go to a medical center where the caregivers used to attach some electrodes to the patient's body and gather the heart signal. Nowadays, with the irruption of the \gls*{IOT} together with the vast and affordable amount of medical sensors that almost everybody has access to, it is reasonably easy to measure the heart signal for many and different purposes. Smartphones \cite{Kwon2012}, smartwatches \cite{AppleWatch}, sport bands \cite{Armstrong2007} or even webcams \cite{Calleja2015} can be used to extract such a signal with no effort.

Among other consequences that both \gls*{IOT} and biometric have brought, is that people can create a so-called \gls*{BAN} by having all the sensors and devices constantly gathering information about themselves. When this \gls*{BAN} is provided with wireless connectivity, we say that the \gls*{BAN} is a \gls*{WBAN}. 

The benefits are unquestionable and measured data can be accessible by anyone, anywhere around the world. We can find examples on remote monitoring of elderly people; athletics trying to improve their performance; to generate a more detailed clinical history without invading the privacy of people, or; to help caregivers to assist people in non-developed countries without the obligation of being physically there \cite{Baig2013,Salayma2017,Movassaghi2014}.

However, the addition of wireless capabilities to the \glspl*{BAN} does not come without problems. New and challenging security and privacy issues mostly associated with the management of the acquired data arise \cite{Salayma2017,Movassaghi2014}. Questions like, how the sensitive information about the user's health is transmitted over the channel; how to gather biometric signals to be used in cryptographic protocols; how to protect those signals from attackers who try to get access to sensitive data; how and where the data are stored; who has access to the data, or; what the purpose of accessing to the data is, are just a few examples of open issues that research community is trying to solve during the last years \cite{Salayma2017,Movassaghi2014} and remain unanswered. 

The heart rate has a chaotic nature \cite{Vibe1995,Lu2003}. The \gls*{ECG} is the only known fiducial-based biological trait used to describe the points of interest which can be extracted from a biological signal \cite{Moosavi2017}. Some examples of fiducial points of the \gls*{ECG} are the P-wave, QRS complex, T-wave, R peaks, or the RR-time-interval. However, it was not until 2003 when Bao \etal\ \cite{Bao2004,Bao2005} proposed to use the \gls*{ECG} as part of biometrics. Concretely, they used the time interval between two consecutive R-peaks (also called \glspl*{IPI}) to generate random nonces to be part of a cryptographic protocol. 

Since then, the researchers have used the \gls*{ECG} signal in many different ways: like a natural \gls*{PRNG} \cite{Camara2018}; as part of an authentication system \cite{Rostami2013}; proposing a key agreement protocol to solving the key distribution issue that symmetric cryptography has \cite{Zaghouani2015,Sammoud2018}; as key generation algorithm \cite{Moosavi2017}; as a cryptographic protocol for biometrics \cite{Vasyltsov2016}, or; like a proof used by sensor nodes to recognize each other \cite{Bao2013}, also known as \gls*{EI} system.


\paragraph{\textbf{Previous Surveys}}
We find authors who focus on concrete applications of cryptographic protocols using biometrics traits like key agreement protocols \cite{Ali2015,Wang2016,Zhao2016}; key generation procedures for \gls*{IOT} \cite{Xu2021}; cryptography in a more general sense \cite{Ribeiro2018}; authentication in \glspl*{WBAN} \cite{Hussain2019,Narwal2021}, and; security and privacy in \gls*{IMD} \cite{Rushanan2014,CAMARA2015272,Yaqoob2019} and wearable devices \cite{Seneviratne2017,Khan2020}. However, most of these works vaguely introduce how the heart signal might be used for biometric but they point out the necessity to a better understanding of how the procedure of transforming the \gls*{ECG} continuous signal to a final cryptographic token works. In our work, we try to fill this gap by collecting and summarizing the contributions published in the literature on biometric protocols based on the heart signal.

Even though many of the aforementioned surveys emphasize the importance of the heart signal in future biometric protocols, we could not find any one covering the heart signal in detail with the only exception of a recent survey published on heart-based biometric authentication protocols \cite{Rathore2020}. In such a survey authors explain in detail how the heart signal is used for authentication purposes. Contrarily to our work, 
authors mainly focus on which machine learning algorithms, distance, and fuzzing algorithms can be used to match two heart signals to be compared and thus, authenticated.  

\paragraph{\textbf{Our contribution}} Our work, aims at complementing the existing surveys on heart-based biometrics. We focus on how the heart signal is transformed to discrete values to be used afterward in security protocols instead of authentication algorithms or the signal acquisition.

We explain and classify the surveyed proposals according to three main parameters: 
\begin{inparaenum}[1)]
    \item the dataset they use for testing their results;   
    \item the delineation algorithms they use to extract the fiducial points, and;
    \item the cryptographic tests they run (if any) to validate how random the extracted signal is.
\end{inparaenum}

We surveyed the most important papers published from 2003 on cryptographic protocols based on heart signals. We restricted ourselves to those proposals published either in journals with Impact Factor\footnote{https://clarivate.com/webofsciencegroup/solutions/journal-citation-reports/} and conferences that listed in the CORE conference ranking\footnote{http://portal.core.edu.au/conf-ranks/}.

\paragraph{\textbf{Organization}} The rest of the paper is organized as follows. We put the heart signal in context, how it is used in biometrics, and the main properties that it has in  \Cref{sec:background}. 
\Cref{sec:data} introduces the main data sources that authors use to test their proposals. In \Cref{sec:delineation}, we detail how the \gls*{ECG} is processed to be used later on cryptographic protocols. The random tests that authors run to validate their proposals can be seen in \Cref{sec:tests}. 
Finally, this paper concludes with some open research issues as well as some conclusions in \Cref{sec:conclusions}. 
\section{Heart-Based Biometrics and Cryptography}\label{sec:background}


Biometrics is defined as the unique physiological or behavioral features of the human body and thus, they can be used to identify persons. On the contrary, Cryptography is an area whose goal is to protect data such that only authorized parties can retrieve such information. 

Nowadays, the primary application of biometrics is authentication and access control \cite{RetinA899930} in office buildings \cite{nieto2002public}, casinos \cite{rowe2006casino,urie2004biometric}, health environments \cite{zuniga2010biometrics, lennox2006combining}, immigration service \cite{van1999illegal}, border control airports \cite{ackleson2005border,adey2009facing}, seaports \cite{ackleson2005border}, border-crossing  \cite{ackleson2005border,cote2008diffuse} and security of in bank systems \cite{von2007biometric, hosseini2012review} among many other examples.

When both biometric and cryptography combine their potential it is possible to create strong cryptographic keys derived from biological signals and therefore establishing a link between that nonce and the person \cite{Hao2006}. One of the most representative examples where biometric and cryptography show their potential when combined is securing \glspl*{WBAN}. 
We can classify biometric systems into \textit{behavioral} and \textit{physical}, according to the nature of the acquired features. In the following, we explain each one of the systems and give examples of proposals where biometric signals are used in cryptographic protocols.

\paragraph{Behavioral} It measures patterns in human activities which make people uniquely identifiable. Some examples of behavioral biometric features are:
\begin{compactenum}
	\item[Gait Analysis.] The way people walk and or using the accelerometer of the smartphone \cite{thang2012gait} can be used to identify people. This is known as gait recognition \cite{zhao20063d}. More information about the phases and cycles of the human walk can be seen in \cite{kharb2011review}.
	
	\item[Keystroke Dynamics.] The goal of this biometric system is to identify persons by monitoring the keyboard inputs and thus generating patterns about the way users type \cite{Monrose1997}. Some applications can be found in \cite{Bergadano2002,Killourhy2009} and some state-of-the-art surveys in \cite{KARNAN2011,teh2013survey}.
	
	\item[Mouse Dynamics.] Mouse distance between two points on the screen, the way the mouse is moved, the drag and drop of the elements or the time that the mouse is idle can be used to identify and authenticate people \cite{feher2012user,shen2013user}. A review performed in \cite{jorgensen2011mouse} summarizes the main contributions performed in this area about mouse dynamics authentication.
	
	\item[Signature Analysis.] This is the most common authentication mechanism  \cite{yeung2004svc2004,doroz2011handwritten} due to devices like touchpads or digital pens \cite{malik2012signature}. An analysis of signature analysis and cryptography was performed by Ballard \etal\ \cite{Ballard2006}.
	
	\item[Voice ID.] This trait is based on multiple voice characteristics such as vocal tracts, mouth, nasal cavities, or lips to authenticate people. Monrose \etal\ \cite{Monrose2002} derived a cryptographic key from the voice. However, despite this it is not a biometric trait that can be used in large and scalable systems \cite{Jain2004}, it is a common technique to be used together with other systems like fingerprint \cite{camlikaya2008multi}.
\end{compactenum}

\paragraph{Physical} It is related to the static traits of a human body that are not subject to change during the time over aging. Some examples of physical biometric features are:

\begin{compactenum}
	\item[Face.] Face trait is a well-known biometric characteristic used nowadays in authentication systems. Eyes, eyebrows, nose, lips, and chin, are commonly used by measuring their spatial relationships to authenticate people \cite{Jain2004}. Chen \etal\ proposed a key generation procedure based on face recognition in 2007 \cite{Chen2007}.
	
	\item[Fingerprint.] This trait refers to a flowing pattern on the fingertip of an individual consisting of ridges and valleys \cite{Jain2004}. It can be acquired by using different technologies as optical sensors or total internal reflection sensors. Fingerprint recognition systems have been incorporated into forensic, civilian, and commercial applications \cite{Marasco:2014}. Recently, Belkhouja \etal\ used the fingerprint together with \glspl*{ECG} as part of a cryptographic protocol for \glspl*{IMD}.
	
	\item[Hand and finger geometry.] These are recognition systems based on measurements taken from the hand, including its shape, size of the palm, and lengths and widths of the fingers \cite{Jain2004}. Handshape biometrics is attractive because it can be captured in a relatively user convenient, shape information requires only low-resolution images \cite{DUTA20092797}.
	
	\item[Heart.] Even though it was thought that the heart signal cannot be used as a biometric trait \cite{Cherukuri2003}, Bao \etal\ \cite{Bao2004} proposed in 2004 one of the first cryptographic protocols based on the \gls*{ECG} signal to secure \glspl*{BAN}. Due to this work revolves around heart-based protocols, we provide more information in further sections. 
	
	\item[Iris.] The complex iris texture carries very distinctive information useful for personal recognition \cite{Jain2004,Wildes1997,ALONSOFERNANDEZ201692}. It contains around 266 visible patterns, which forms the basis of several recognition algorithms. Even on the same person, left and right irises are different but they are unique to an individual and are stable with age \cite{GANESHAN20061}. Examples of cryptographic protocols using the Iris can be seen in \cite{Zheng2006,Hao2006}.
	
	\item[Retina.] A retina-based biometric involves analyzing the layer of blood vessels situated at the back of the eye. It is claimed to be the most secure biometric since it is not easy to change or replicate the retinal vasculature \cite{Jain2004}. Retinal scanning can be quite accurate but does require the user to look into a receptacle and focus on a given point. 
\end{compactenum}


We can categorize biometrics solutions as fiducial, and non-fiducial approaches \cite{Rathore2020}.   The term \say{fiducial points} in biometrics usually refers to the representative characteristics used to characterize the involved signal. However, the process of extracting these fiducial points is sometimes challenging—biological signals, in general, are usually a response to dynamic changes in the organs' work. Thus,  we can detect alterations in the signal in the form of time-varying and non-stationary patterns. When sensors record biometric signals, the measurements can be altered by drift and interferences due to bioelectric phenomenon, errors associated with the sensor itself, noise from electrode-skin contact, and low-frequency interferences, among others \cite{Sayadi2008}.

\subsection{Heart Signal}\label{sec:heart}

\begin{figure}
\centering
\includegraphics[width=\columnwidth]{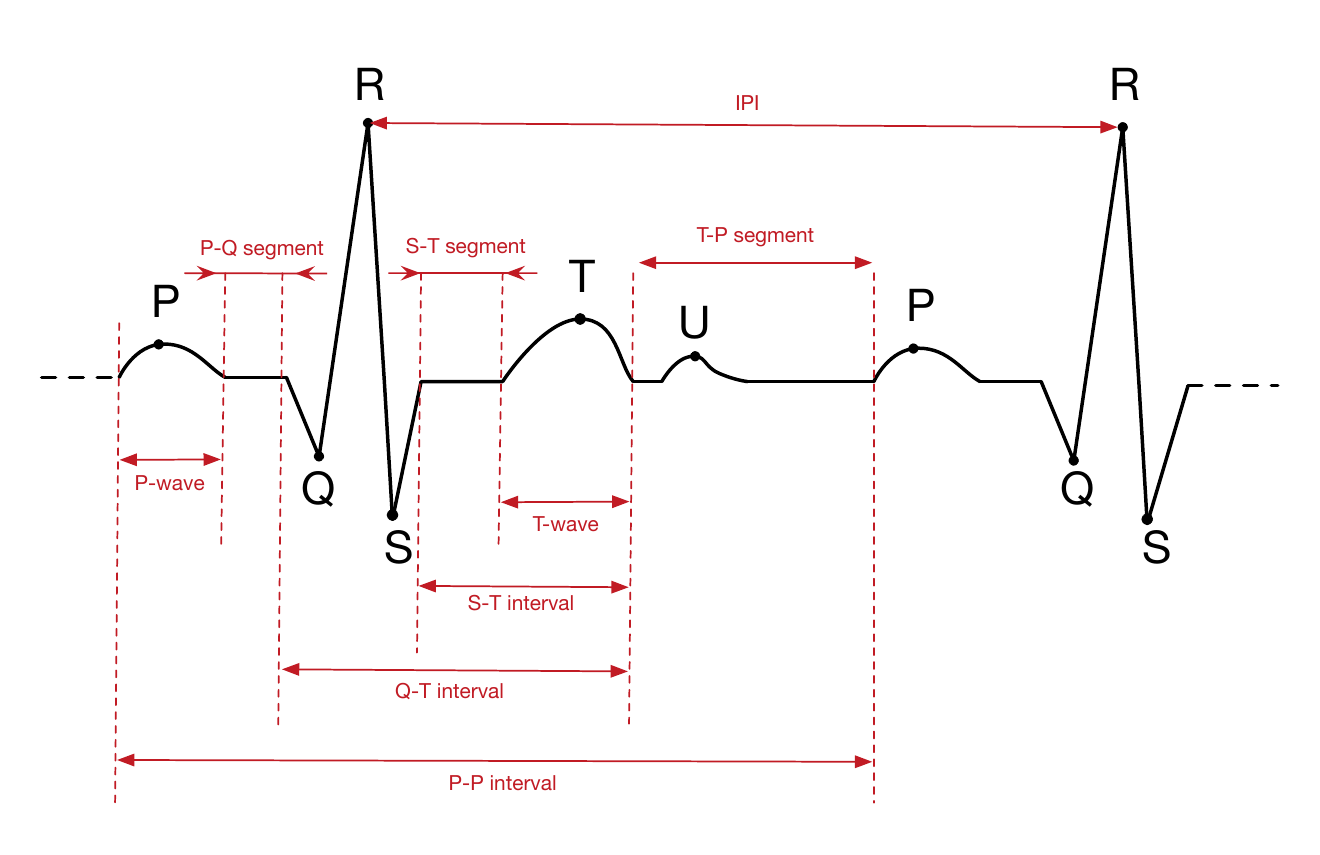}
\caption{A typical Electrocardiogram (ECG) signal and its main features: peaks (P, Q, R, S, T, U), waves, segments and intervals \cite{Ortiz2018}.}
\label{figure:ecg}
\end{figure}

The \gls*{ECG} is the electrical representation of the heart activity. In \Cref{figure:ecg}, we can see the typical shape of an \gls*{ECG} signal as well as the most representative characteristics of it, \eg\ peaks of the waves like P, Q, R, S, T, and U as well as the segments and the intervals---which is the time when the waveform starts and ends. Among all the characteristics of the \gls*{ECG}, the QRS complex is the most important one \cite{Martinez2004} that starts when the Q wave begins and ends when the S waveform finishes.

Rathore \etal\ \cite{Rathore2020} described in detail how the heart signal works as well as the most important sensors that can be used to gather it. In the following, we explain the main concepts that cryptographic protocols based on the heart signal use: fiducial points, heart-rate variability as well as which ones are the most used sensors in the literature.

\subsubsection{Gathering the Heart Signal} Nowadays, 
we can find rings \cite{rings}, watches \cite{AppleWatch,galaxy} and wrist bands among many other devices with heart-sensing capabilities. Even though some of them present real challenges to the research community giving the inaccuracy \cite{Colvonen2020}, some others have helped users to detect heart diseases \cite{AppleDetection} or even helping in the COVID-19 early detection due to the variation of the \gls*{HRV} \cite{AppleCovid}.

To gather the heart signal, we can use sensors like the \gls*{BP}, \gls*{ECG} and \gls*{PPG}. \gls*{BP} sensor measures the blood pressure in the systemic circulation and the heartbeats are reflected in the up and down fluctuation of the arterial pressure. To gather the \gls*{ECG} signal, usually, some electrodes are placed on the chest of the patient to detect the electrical changes in the heart and hence, generate the signal. The \gls*{PPG} detects the pulse of the heart by measuring the amount of light reflected in the skin to a photodiode. 

\subsubsection{Fiducial Points in ECG Signal} The heart signal is not exempt from the problems that gathering biological signals has. When sensing the heart signal, apart from the technical difficulties due to, for instance, the noise of the signals; missed data during the gathering phase; delays, or; the bearer's movements \cite{Seepers2014}, there are two main aspects to consider when using the heart signal in cryptographic protocols:
\begin{inparaenum}[1)]
    \item external alterations: it has been proven that parameters such as \gls*{HRV} can alter the peak detection algorithm \cite{Lin2014}, and;
    \item length of the signal: it was thought that sensors have to gather for about 30 seconds \cite{Rostami2013,Bao2008,Xu2011,Moosavi2017} to generate 128 bits token. However, a recent study \cite{Ortiz2019} increased that time to about 60 seconds to generate the same length token.
\end{inparaenum}

Concretely, some examples of fiducial points usually used in \gls*{ECG}-based cryptographic protocols are the peaks of the signal, \ie\ P, Q, R, S, T, and U (see \Cref{figure:ecg}). Indeed, one of the most common fiducial points used in the literature is the R-peak which, by using the Pan-Tomkins's QRS detection algorithm \cite{pan1985real}, achieves 99\% of accuracy \cite{Lee2018}, probably because is the easiest to detect, being detectable by most of the sensors in the market. However, any point in the continuous function of the signal can also be taken as a fiducial point despite the detection accuracy is not as good as the R-peak extraction and is still challenging nowadays \cite{Lee2018}.

In general, we can distinguish between extracting fiducial points using the time or the frequency domains of the signal. This last procedure might be used independently, \ie\ extracting the fiducial points of the frequency domain  \cite{Camara2018,Zheng2016,yao2011biometric,Venkatasubramanian2010,Venkatasubramanian2010b}, or as a complement to the information extracted from the time domain  \cite{Moosavi2017,Moosavi2017b,Hu2013,Xu2011}. 

Remarkably, the number of fiducial points that can be extracted from the \gls*{ECG} signal depends on the frequency. For instance, if a signal is gathered at 500Hz, it means that a new measurement is read each 2 milliseconds, \ie\ $t=\frac{1}{500}=2\times10^{-3}$s. We include the frequency of the signals of the most common databases used in the literature in \Cref{table:thew_description,table:physionet_description}.

We present the most common methods used by extract fiducial points from the heart signal, called \textit{delineation algorithms} in \Cref{sec:delineation}, where we also classify all the surveyed papers according to the algorithm they use.

\subsubsection{Heart-Rate Variability}

The \gls*{HRV} is the variance in time between heartbeats. These heartbeats are essentially controlled by both the sympathetic and the parasympathetic nervous systems \cite{Acharya2006}. Roughly speaking, when we say that a heart beats 60 times per minute, it does not mean that there is a heartbeat each second.

In 2001 and 2003, researchers \cite{Murthy2001,Teng2003} analyzed the correlation of the time interval between R-peaks, also known as \glsentryfull{IPI}, of the \gls*{ECG}. Later, in 2006, Acharya \etal\ \cite{Acharya2006} analyzed how the R-peaks are related to \gls*{HRV}. These works concluded that the time between two consecutive R-peaks and the \gls*{HRV} can be considered to be similar when the patient is resting. Similarly, after doing exercise neither the R-R nor the P-P intervals can be considered similar to the \gls*{HRV}.

\subsection{Peak Misdetection Recovery Approach}
It has been stated in different works that the fiducial points, and more concretely, the \glspl*{IPI} gathered from the heart signals by the sensors actually differ from one sensor to others due to the bearer's movements, errors during the gathering phase or just because of the noise in the signal \cite{Seepers2014}. In conclusion, the so-called peak misdetection negatively affects the generated token because sensors actually get different tokens and thus, cannot be used as a secret key \cite{Kim2018}.  

Researchers proposed solutions for both detection and recovery actions. For instance, in \cite{Kim2018} authors propose a method to detect those missed peaks as well as a recovery key exchange protocol. The authors modeled the heart behavior in such a way that a new R-peak should come each 600--1200 ms. Hence, if a new peak comes after that threshold the system detects that a peak has not been detected. 

Most of the aforementioned works rely on coding theory to recover errors in the transmission of the signal. In particular, they rely on \gls*{BCH} codes to \textit{synchronize} the heart signal in different sensors. However, Ortiz \etal\ \cite{Ortiz2019} demonstrate that just using \gls*{BCH} algorithms are not enough to synchronize the signal. To do so, the authors used over 20 datasets from the Physionet repository with the signal being gathered from, at least, two devices. They conclude that a preprocessing of the signal is needed before the \gls*{BCH} can recover the missing bits with some confidence level.

\subsection{Fuzzy Vault, Fuzzy Commitment, and Fuzzy Extractor}

Both fuzzy vault and fuzzy commitment schemes were proposed by Juels \etal\ \cite{Juels1999,Juels2002}. In \emph{fuzzy vault schemes} \cite{Juels2002}, Alice creates a secret and locks it in a structure called \textit{vault}. The secret can be unlocked if and only if the vault created by Bob overlaps substantially with the one previously created by Alice. On the contrary, in \emph{fuzzy commitment schemes} \cite{Juels1999},
a cryptographic key is hidden using Alice's biometric data and transmitted later on to Bob. When Bob receives the message, in the decommitment phase he reveals the key if and only if Bob's biometric trait is close enough to Alice's one. Even though both heavily rely on the \glspl*{ECC}, fuzzy vault schemes are based on polynomial calculation and they also implement a computationally demanding reconstruction operation. Therefore,  fuzzy commitment schemes are recommended for high-constrained \glspl*{WBAN} sensors \cite{Zheng2015Fuzzy}.  

\begin{figure}
\centering
\includegraphics[width=8.5cm]{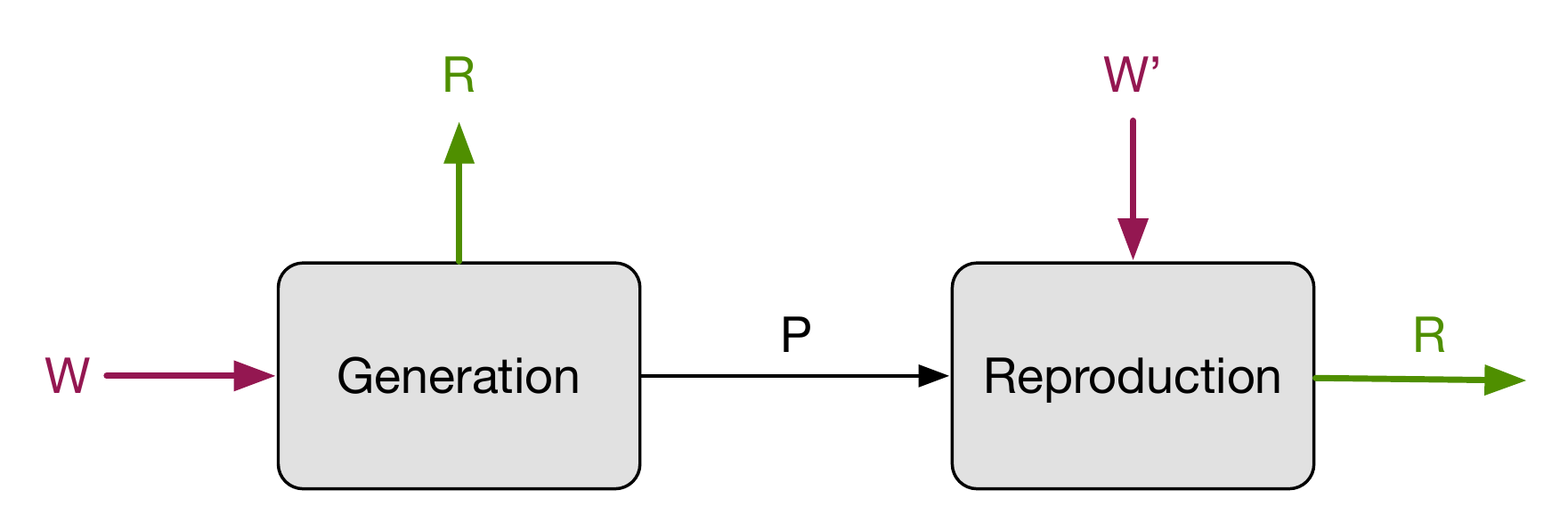}
\caption{Fuzzy extractor architecture \cite{Natgunanathan2016}.}
\label{fig:fuzzy_extractor}
\end{figure}

In 2004, Dodis \etal\ \cite{Dodis2004} proposed a more general construction of the aforementioned schemes called \textit{fuzzy extractor}. Formally, a fuzzy extractor is a function $f$ that uses as input a biometric signal $w$ and produces both a random string $R$ and a public parameter $P$. Fuzzy extractors are particularly suitable for cryptographic protocols because when the input $w'$ changes slightly, \ie\ $w'= w + \epsilon$ for a very small $\epsilon$, the random output $R$ remains invariant.

A fuzzy extractor is composed of a \emph{generation} and a \emph{reproduction} phases \cite{Dodis2004}. In the generation phase, a biometric signal $w$ is used as input to generate a secret $R$ and a public $P$ values (see \Cref{fig:fuzzy_extractor}). In the reproduction phase, a fresh biometric signal $w'$ is provided as input together with the public parameter $P$. If and only if the distance between these two biometric values---typically the Hamming distance---is less than a given threshold $t_r$ ($Hamming(w,w')<t_r$), then $R$ is retrieved. 

Although fuzzy extractors are the most appropriate construction for extracting biological traits to be used in cryptographic protocols \cite{Marin2019}, they have performance issues due to the complexity of the computations. A solution is using Bose–Chaudhuri–Hocquenghem (BCH) codes, a generalization of Hamming codes for multiple error correction \cite{Hocquenghem1959,Bose1960}. 

Formally, for any positive integer $m\ge3$ and $t<2^m-1$ there exists a binary BCH code that:
\begin{compactitem}
    \item Block length: $n=2^m-1$;
    \item Number of parity checks is $n-k\le mt$ and;
    \item Minimum distance: $d_{min}\ge 2t+1$.
\end{compactitem}

We call this code a \emph{t-error-correcting} BCH code where the generator polynomial $g(X)$ of the of length $2^m -1$ is the lowest-degree polynomial over $GF(2)$ with $\alpha, \alpha^2, \cdots, \alpha^{2t}$ as roots. Let $\Phi_i(X)$ be the minimal polynomial of $\alpha^i$, then $g(X)$ is the Least Common Multiple (LCM) of $\Phi_{1}(X), \Phi_{2}(X),.., \Phi_{2t}(X)$. The degree of $g(X)$ is at most $mt$, \ie\ the number of parity-check digits ($n-k$) of the code is at most $mt$.

Block codes are implemented as $(n,k,t)$ codes where $n$ is the codeword, $k$ the original information bits and $t$ is the correction capability, \ie\ this BCH code is capable of correcting any combination of $t$ or fewer errors in a block of $n = 2^m$ — $1$ bits. 

As an example, given BCH(63, 5, 2) where $n=63$, $k=5$, and $t=2$, then the generator polynomial $g(X)$ is $g(X)=(1+x^3 +x^4 +x^5 +x^9 +x^{10} +x^{12})$. In this example, the \gls*{BCH} can recover at most 2 different bits from words of 64 bits (see \cite{Bose1960,Hocquenghem1959,BCH} for a more detailed description of \gls*{BCH} codes). 

\begin{figure}
\centering
\includegraphics[width=9.15cm]{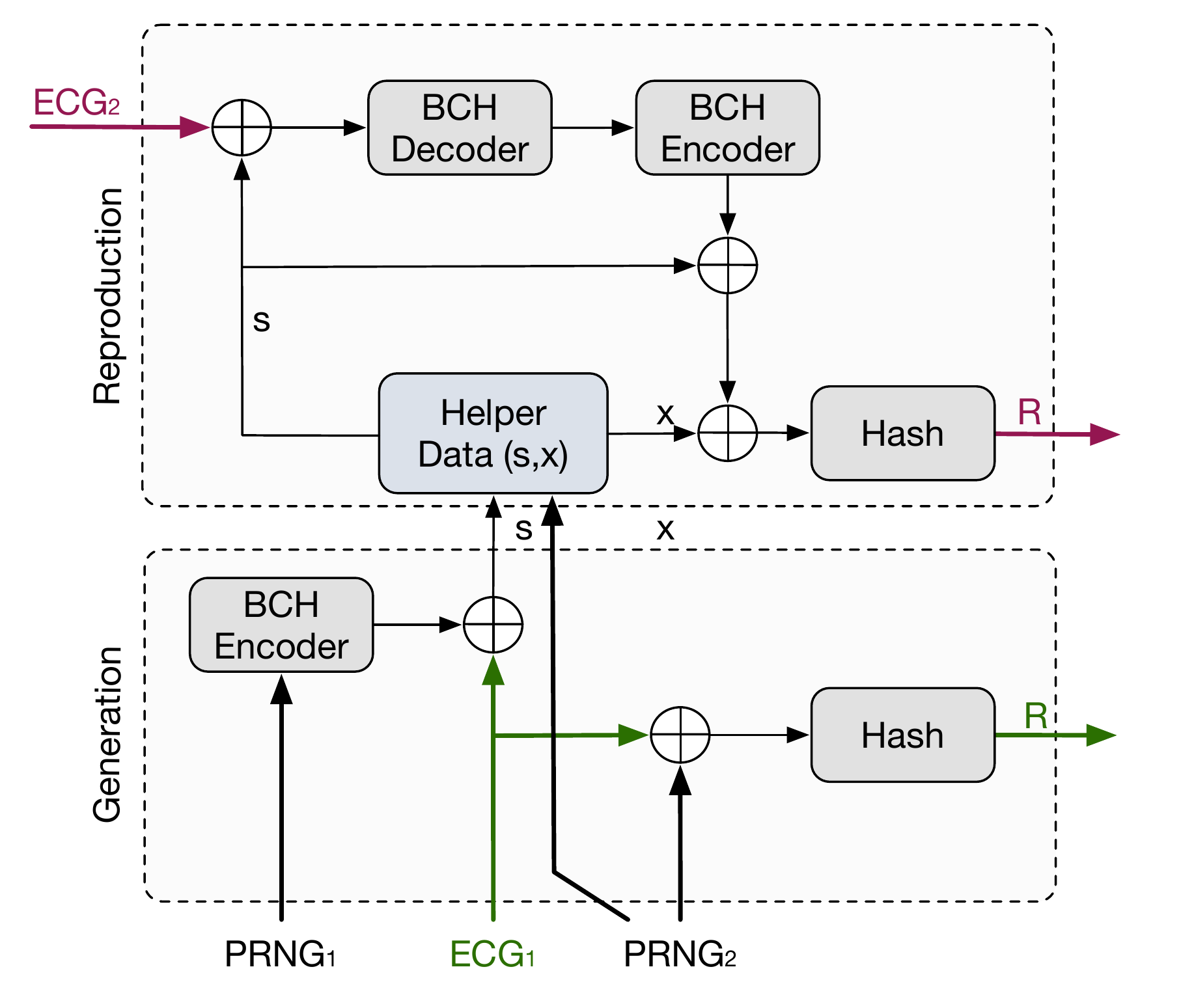}
\caption{Fuzzy extractor \cite{Ortiz2019}.}
\label{figure:BCH}
\end{figure}

\subsubsection{Heart Signal, Fuzzy Extractors, and BCH Codes}
When applied to the heart signal, \gls*{BCH} codes have been widely used to recover the same signal gathered by multiple sensors located in different parts of the body.

Sun \etal\ \cite{Sun2017} proposed a fuzzy commitment-based key distribution scheme to distribute the keys among the sensor nodes, and they used BCH(127,15,27) configuration, meaning that the error correction is set to over 21\%.

Sammoud \etal\ \cite{Sammoud2018} used \gls*{BCH} to synchronize somehow the same heart signal gathered from two sensors at the same time. In more detail, they used BCH(31,6,3) as an error-correcting code (over 10\% of the errors) and claimed to achieve 100\% recoverability of the generated keys.

Ortiz \etal\ \cite{Ortiz2019} implemented a fuzzy extractor based on \gls*{BCH} codes (see \Cref{figure:BCH}). They used BCH(127,50,13),  meaning that they could recover 10\% of the errors during the transmission of the heart signal among devices. Their goal is to empirically demonstrate that the heart signal should be synchronized before the \gls*{BCH} algorithm can recover the same nonce on different devices. If the signal is not synchronized, the derived tokens might differ up to the point that \gls*{BCH} cannot recover the errors. 


\section{ECG Public Repositories}\label{sec:data}

In this section, we detail how the surveyed papers test their proposals with respect to the heart signals. In particular, we found two main groups, authors who: 
\begin{inparaenum}[i)]
    \item use proprietary data, and;
    \item use data from public repositories. 
\end{inparaenum}

There are three main public repositories that authors use:  Telemetric and \gls*{ECG} Holter Warehouse Project (THEW) \cite{Couderc2010},  Biosec \cite{BioSec}, and;  Physionet \cite{PhysioNet} datasets. Due to the number of researchers who use Physionet in comparison to the other two mentioned databases, we are significantly giving more details about it, \ie\ we explain its composition and how the data should be extracted and processed. 

\begin{table*}[th]
\caption{Summary of the THEW databases. SID refers to the unique ID in the repository; Leads refers to the number of leads acquiring in most recordings; Sampling refers to the sampling frequency of the ECG waveforms; ECGs is the number of ECG recordings; Patients is the number of subjects involved, and; Size is the storage space of the database.}
\label{table:thew_description}

\centering
\resizebox{\textwidth}{!}{%
\begin{tabular}{l l r r r r r r }
\toprule
   \textbf{Database} & \textbf{SID} & \textbf{Leads} & \textbf{Sampling} & \textbf{\#ECGs} & \textbf{\#Patients} & \textbf{Size}\\ 
\midrule
Acute Myocardial Infarction	& E-HOL-03-0160-001	& 3	& 200Hz & 160 & 93	& 15.2GB \\
Coronary Artery Disease & E-HOL-03-0271-002 & 3 & 200Hz & 271 & 271 & 26.2GB \\
Healthy & E-HOL-03-0202-003 & 3 & 200Hz & 202 & 202 & 19.2GB \\
Thorough QT study \#1 & E-HOL-03-0102-005 & 3 & 200Hz & 102 & 34 & 4.5GB \\
Thorough QT study \#2 & E-HOL-12-0140-008 & 12 & 1kHz & 140 & 70 & 267GB \\
Torsades de Pointes (TdPs) & E-OTH-12-0006-009 & 12 & 180Hz & 6 & 6 & 1.3GB \\
Sotalol IV and History of TdPs & E-OTH-12-0068-010 & 12 & 1kHz & 68 & 34 & 244MB \\
AF and cardioversion & E-OTH-12-0073-011 & 12 & 1kHz & 73 & 73 & 1.7GB \\
Chest Pain (IMMEDIATE LR ECG) & E-HOL-12-1172-012 & 12 & 180Hz & 1,172 & 1,154 & 338GB \\
Genotyped Long QT syndrome & E-HOL-03-0480-013 & 2 or 3 & 200Hz & 480 & 307 & 43.2GB \\
Chest Pain (IMMEDIATE HR ECG) & E-HOL-12-0171-014 & 12 & 1kHz & 171 & 171 & 296GB \\
Exercise testing and perfusion imaging & E-OTH-12-0927-015 & 12 & 1kHz & 927 & 927 & 23GB \\
ESRD patients during and after hemodialysis & E-HOL-12-0051-016 & 12 & 1kHz & 51 & 51 & 187GB \\
FDA1- quinidine, verapamil, ranolazine, dofetilide & E-OTH-12-5232-020 & 12 & 1kHz & 5,232 & 22 & 1.7GB \\
FDA2- quinidine, verapamil, ranolazine, dofetilide & E-HOL-12-0109-021 & 12 & 1kHz & 109 & 22 & 231GB \\
AF Conversion & E-OTH-12-0089-022 & 12 & 1kHz & 26 & 26 & 3.23GB \\
\say{Strict} LBBB & E-OTH-12-0602-024 & 12 & 1kHz & 602 & 602 & 157MB \\
Young Healthy & E-OTH-12-0689-025 & 12 & 500kHz & 689 & 689 & 207MB \\
\multicolumn{7}{l}{\textbf{Collaborative studies (require the submission of a research proposal to an ad-hoc THEW committee})} \\
DEFINITE Study (NorthWestern Univ.) & E-HOL-03-0401-017 & 3 & 500Hz & 401 & 236 & 110GB \\
Occluded Artery Trial (Stony Brooke Univ.) & E-OTH-03-0802-018 & 3 & 500Hz & 802 & 223 & 6GB \\
Quinidine (AZCERT) & E-OTH-12-2365-019 & 12 & 500Hz & 2423 & 24 & 17MB \\
IQ-CSRC & E-HOL-12-0118-023 & 12 & 1kHz & 118 & 20 & 22.8MB \\
\textbf{TOTAL} &  &  &  &  &  &  \textbf{15TB} \\
\bottomrule
\end{tabular}%
}
\end{table*}

\begin{table}[t]
\caption{Summary of the Pysionet databases. SID refers to the unique ID in the repository; Leads refers to the number of leads acquiring in the most recordings; Sampling refers to the sampling frequency of the ECG waveforms; ECGs is the number of ECG recordings.}
\label{table:physionet_description}
\centering
\resizebox{\columnwidth}{!}{%
\begin{tabular}{l l r c r }

\toprule
   \textbf{Database} & \textbf{SID}  & \textbf{Leads} & \textbf{Sampling} & \textbf{\#ECGs}\\ 
\midrule
  Tachycardia & aami-ec13 \cite{aami}  & 1 & 720Hz & 10\\ 
  Atrial fibrillation & afdb \cite{afdb} & $\ge$2 & 0.1Hz $\sim$ 40Hz & 23\\ 
  Paroxysmal atrial fibrillation & afpdb \cite{afpdb} & $\ge$2 & 128Hz & 300\\ 
  Healthy and ventricular ectopy & ahadb \cite{ahadb}  & $\ge$2 & 250Hz & 2\\ 
  Tachycardia & apnea-ecg \cite{apnea-ecg} & 1 & 100Hz & 77\\ 
  Healthy & cebsdb \cite{cebsdb} & $\ge$2 & 5,000Hz & 60\\ 
  Holter recordings & cdb \cite{cdb} & --- & 250Hz & 53\\
  Ventricular problems & cudb \cite{cudb}  & --- & 250Hz & 9\\
  No info is provided & ECG-ID \cite{ecgid} & 1 & 500Hz & 90\\
  Myocardial and hypertension & edb \cite{edb} & $\ge$2 & 250Hz & 90\\
  Healthy & fantasia \cite{fantasia} & 1 & 250Hz & 40\\ 
  Atrial fibrillation or flutter & iafdb \cite{iafdb} & $\ge$2 & 1kHz & 32\\ 
  Coronary artery disease & incartdb \cite{incartdb} & $\ge$2 & 257Hz & 75\\ 
  Paroxysmal & ltafdb \cite{ltafdb} & $\ge$2 & 128Hz & 84\\ 
  \gls*{ICU} & mimic2wdb \cite{mimic2} & $\ge$2 & 125Hz & 25,328\\ 
  Arrhythmia & mitdb \cite{mitdb} & $\ge$2 & 360Hz & 48\\
  Unstable patients in \gls*{ICU} & mghdb \cite{mghdb}  & 3 & 360Hz & 202\\ 
  No significant arrhythmias & nsrdb \cite{nsrdb} & $\ge$2 & 128Hz & 18\\ 
  Mitdb with noise & nstdb \cite{nstdb} & $\ge$2 & 360Hz & 15\\ 
  Healthy & prcp \cite{nstdb} & $\ge$2 & 250Hz & 10\\ 
  Myocardial and Healthy controls & ptbdb \cite{ptbdb} & 14 & 1kHz  & 545\\ 
  Holter recordings & qtdb \cite{qtdb} & $\ge$2 & 250Hz & 105\\ 
  Arrhythmia & sddb \cite{sddb} & $\ge$2 & 250Hz & 22\\ 
  Hypertension & shareedb \cite{shareedb} & $\ge$2 & 128Hz & 139\\ 
  Sleep apnea syndrome & slpdb \cite{slpdb} & $\ge$2 & 250Hz & 18\\
  Stress tests & stdb \cite{stdb} & 2 & 360Hz  & 28\\ 
  Partial epilepsy & svdb \cite{svdb}  & $\ge$2 & 128Hz& 70\\ 
  Partial epilepsy & szdb \cite{szdb}  & 1 & 200Hz & 7\\ 
  Myocardial problems & twadb \cite{twadb}  & $\ge$2 & 500Hz & 100\\ 
  Tachycardia & vfdb \cite{vfdb} & $\ge$2 & 250Hz & 22\\ 
\bottomrule
\end{tabular}%
}
\end{table}

\paragraph{\textbf{BioSec}} Biosec is a repository composed by 7 databases maintained by researchers from Toronto University\footnote{\url{https://www.comm.utoronto.ca/~biometrics/databases.html}}.  It essentially has databases of healthy people. However, the specifications of this repository are in contradiction in the official website so we decided to not include them in this study.

\paragraph{\textbf{THEW}} Thew is a public repository created and maintained by the University of Rochester Medical Center \footnote{\url{http://thew-project.org/index.htm}}. It is composed of 22 databases going from healthy to people with heart or renal diseases (see \Cref{table:thew_description} for a complete description of the repository). What makes this database particularly interesting is the number of healthy people that it has. concretely, this repository has two databases with \glspl*{ECG} named \say{E-OTH-12-0689-025} with 689 and \say{E-HOL-03-0202-003} and 202 healthy patients. In most of the databases of this repository, files are given in both \cpp\ and \m\ formats to be managed under C++ and Matlab respectively. Additionally, authors have published some other features to read the \gls*{ECG} and the annotation files in Matlab directly from their server\footnote{More info at \url{http://thew-project.org/THEWFileFormat.htm}}.

\paragraph{\textbf{Physionet}} This is a public repository constantly updated by medical researchers who share sensitive information about patients. It provides an open-source software named PhysioToolkit which can be used to read and display these signals. At the time of writing, Physionet has more than 75 databases classified into two main families: clinical databases (include demographics, vital sign measurements made at the bedside, laboratory test results, procedures, medications, caregiver notes, images and imaging reports, and mortality) and waveform databases (high-resolution continuous recordings of physiological signal). Each family has different categories: biomedical, brain, or cardiopulmonary signals. Additionally, the health condition of the patients varies considerably: healthy people, heart diseases, apnea, or epilepsy among others.

\noindent Focusing on the structure of the files, we can have five main different files:
\begin{inparaenum} [1)]
    \item header files (\hea). These files contain the metadata of the record.
    \item signal annotations (\atr). These files contain the annotations of the biometrical data;
    \item biometrical data (\dat). These files have all the gathered personal information of the patient;
    \item \path{RECORDS} files where all the names of the files are listed, and;
    \item \path{ANNOTATORS} files where the information of how to read the \atr\ files is explained.
\end{inparaenum}

\subsection{Summary of Surveyed Papers}

\begin{table}[t!]
\caption{Description about the composition of the dataset(s) taken by authors with the number of records in parenthesis. Patients is the total number of patients taken for the experiments. Private indicates if private data were used. Description includes detailed information provided by the authors.}
\label{table:db_authors}
\begin{threeparttable}
\centering
\resizebox{\columnwidth}{!}{%
\begin{tabular}{l r c p{5cm} }
\toprule
    & \textbf{\#Patients} & \textbf{Private} & \textbf{Description}\\ 
\midrule
  Altop \etal\ \cite{Altop2017} & 50 & &mimic II waveform \\
  Bai \etal\ \cite{Bai2019} & 11 & & mitdb\\
  Bao \etal\ \cite{Bao2004} & 4 & \checkmark &\\
  Bao \etal\ \cite{Bao2005} & 12 & \checkmark &\\
  Bao \etal\ \cite{Bao2008} & 99 & \checkmark &\\
  Bao \etal\ \cite{Bao2013} & 14 & \checkmark &\\
  Belkhouja \etal\ \cite{Belkhouja2019} & --- & --- &\\
  Camara \etal\ \cite{Camara2018}& 202+ & \checkmark & 202 from E-HOL-03-0202-003 \\
  Chen \etal\ \cite{Chen2012} & 6 & \checkmark &\\
  González-Manzano \etal\ \cite{Manzano2017} & 199 &  & E-HOL-03- 0202-003\\
  Hong \etal\ \cite{Hong2011} & 10 & \checkmark &\\
  Hu \etal\ \cite{Hu2013} & 11 & &Physionet\tnote{1}\\
  Ingale \etal\ \cite{Ingale2020} & 1,691 &  & ptb (290); mitdb (47); cebsdb (20); ECG-ID (90), CYBHi (125), ECG-BG (1,119)\\
  Karthikeyan \etal\ \cite{Karthikeyan2018}& 70 &   & mitdb (47), and afdb (23) \\
  Kim \etal\ \cite{Kim2018}& --- & &PhysioNet\tnote{1}\\ 
  Koya \etal\ \cite{Koya2018} & --- & &ptbdb and mitdb\\
  Lin \etal\ \cite{Lin2019} & 23 & \checkmark & 7 females and 16 males \\
  Mahendran \etal\ \cite{Mahendran2020} & --- & --- &\\
  Moosavi \etal\ \cite{Moosavi2017}& 15 & &mitdb\\
  Moosavi \etal\ \cite{Moosavi2017b}& 239 & &ptdb\\
  Pirbhulal \etal\ \cite{Pirbhulal2015} & 20 &  & nsrdb and ltdb\\ 
  Pirbhulal \etal\ \cite{Pirbhulal2018} & 89 & \checkmark & mitdb (20), and a private dataset (25 healthy and 44 with cardiac diseases)\\ 
  Rostami \etal\ \cite{Rostami2013}& 587 & & mitdb (47); ptbdb (290), and; mghdb (250)\\
  Peter \etal\ \cite{Peter2016} & 4 & \checkmark &authors of the paper\\
  Poon \etal\ \cite{Poon2006} & 99 & \checkmark &\\
  Reshan \etal\ \cite{Reshan2019} & --- & & nsrdb, mitdb, and edb\\
  Sammoud \etal\ \cite{Sammoud2018} & 15 & --- & ptb (15)\\
  Seepers \etal\ \cite{Seepers2015}& 48 & & mitdb and fantasia database\\
  Seepers \etal\ \cite{Seepers2017}& 153 & & 42 from mitdb and fantasia, and 111 from BioSec dataset\\
  Sun \etal\ \cite{Sun2017} & 5 & \checkmark & \\
  Vasyltsov \etal\ \cite{Vasyltsov2016} & --- &  & mitdb \\ 
  Venkatasubramanian \etal\ \cite{Venkatasubramanian2010,Venkatasubramanian2010b} & 10 & \checkmark & mimic and 10 from a private dataset\\
  Wu \etal\ \cite{Wu2018} & 126 & \checkmark & private dataset (74), and Physionet\tnote{1} (52). \\
  Xu \etal\ \cite{Xu2011} & --- & &Physionet\tnote{1}\\ 
  Yao \etal\ \cite{yao2011biometric} & 294 & &qtdb\tnote{2}\\
  Zaghouani \etal\ \cite{Zaghouani2015} & 84 & &nsrdb (36) and mitdb (48)\\
  Zaghouani \etal\ \cite{Zaghouani2017} & 90 &  &ECG-ID\\
  Zaghouani \etal\ \cite{Zaghouani2017b} & 290 &  &ptbdb\\
  Zhang \etal\ \cite{Zhang2012} & 84 & \checkmark & edb (64) and a private dataset (20)\\ 
  Zhang \etal\ \cite{Zhang2010} & 79  &  & from stdb \tnote{3}\\ 
  Zheng \etal\ \cite{Zheng2015} & 167 & &nsrdb (18); edb (79); mitdb (47), and; afdb (23)\\ 
  Zheng \etal\ \cite{Zheng2016,Zheng2019}& 97 & &nsrdb (18), and edb (79)\\ 
  \bottomrule
\end{tabular}%
}
\begin{tablenotes}
\item[1] \scriptsize It is not specified in the paper
\item[2] \scriptsize Authors claim they use 294 patients but this qtdb only consists of 105 patients
\item[3] \scriptsize Authors claim they use 79 patients but this stdb only consists of 28 patients
\end{tablenotes}
\end{threeparttable}
\end{table}

We include in \Cref{table:physionet_description} all the databases we found in the literature together with the description that Physionet provides for each dataset (column 1); the short name by which is known (column 2); if more than one \gls*{ECG} channel is provided (column 3); the frequency of sampling (column 4), and; the number of files that each one of the databases is composed of (column 5).

Given the amount of data provided by Physionet, it is so popular among researchers. However, we found that authors use arbitrary databases in their proposals. Since there is no formal framework, rules, or tests that authors can run to compare their proposals with others, it is hard to objectively say which proposal solves what problem. Just to cite a particular example, Zheng \etal\ \cite{Zheng2016} compare their proposal with Zhang \etal's work \cite{Zhang2012} in terms of the \gls*{NIST} performance. However, this comparison is far from being objective since authors use not only different samples but also different random tests. 

In \Cref{table:db_authors}, we include a summary of the surveyed proposals (column 1); the number of patients involved in the experiments (column 2); whether they use or not private data (column 3), and; some description about the composition of the dataset when it is possible. Note that there are some cases \cite{Camara2018,Pirbhulal2018,Venkatasubramanian2010,Venkatasubramanian2010b,Zhang2012} where authors apart from using private dataset they also use information from public repositories. 

Regarding the condition of the patients, the authors consider neither the frequency of the acquired signal nor the cardiac disease of the patients. For instance, Rostami \etal\ \cite{Rostami2013} tested their proposals against patients suffering from myocardial and healthy people (ptbdb), arrhythmia (mitdb) and unstable patients in \gls*{ICU} (mghdb) with frequency of sampling of 1kHz, 360Hz and 360Hz respectively---note that this is an example but there are many others \cite{Zaghouani2015,Zhang2012,Koya2018,Karthikeyan2018,Seepers2017,Seepers2015,Zheng2016,Zheng2015,Zheng2019}. Apart from that, there are three special cases where authors did not mention which databases of the Physionet repository they used to validate their experiments \cite{Hu2013,Kim2018,Xu2011}, or those that exclusively use private datasets \cite{Lin2019,Bao2004,Bao2005,Bao2008,Bao2013,Hong2011,Poon2006,Chen2012} or a combination of private and public datasets \cite{Camara2018,Pirbhulal2018,Peter2016,Venkatasubramanian2010,Venkatasubramanian2010b,Zhang2012}.

\section{ECG Fiducial Points Delineation Algorithms}\label{sec:delineation}

Since most of the clinical information that the \gls*{ECG} has is in the intervals and the amplitudes of the wave peaks and boundaries, an algorithm must be used to extract such information from the \gls*{ECG}. This procedure is known as delineation and it is still challenging nowadays \cite{Martinez2004}. 

The first step to extract the fiducial points is to get an R-peak \cite{Kohler2002}. After that, the QRS complex and the P and T waves can be delineated. From there, forward and backward seek windows can be defined as well as other techniques to enhance the needed waves and fiducial points can be applied \cite{Martinez2010}. However, as Martinez \etal\ \cite{Martinez2004} pointed out, the detection of the wave onset and offset directly from the \gls*{ECG} is not trivial due to the signal amplitude is significantly low in comparison to the wave boundaries. In addition to that, the noise level can also be higher than the signal itself.

Although there are many ways to extract fiducial points from the heart signal, \eg\ mathematical models \cite{Madeiro2013}, \gls*{FFT} \cite{Venkatasubramanian2010}, \glspl*{WT} \cite{Akhbari2016,Camara2018,Rincon2011}, hidden Markov models \cite{Akhbari2016,Graja2005}, or artificial intelligence algorithms \cite{Lin2010,Mehta2010,Saini2013}, in heart-based cryptographic protocols we could exclusively find authors who use:
\begin{inparaenum}[i)]
    \item time-domain algorithms (\gls*{IPI}-based protocols);
    \item frequency-domain protocols (Fourier-based protocols), and;
    \item time-frequency-domain protocols (Wavelet Transforms-based protocols).
\end{inparaenum}  

To extract the fiducial points from the \gls*{ECG} signal, we can either use the time and the frequency domains of the \gls*{ECG} signal or a combination of both. The choice basically depends on the deployment scenario. The former is not computationally demanding and highly constrained devices can process the \gls*{ECG}, compute the R-peaks, and perform small computations. On the contrary, the execution of delineation algorithms based on \glspl*{WT} or \gls*{FFT} require much more complex operations which directly affects the energy consumption.

\paragraph{\textbf{Quantization Algorithms}} Once the peaks have been extracted, some authors use the term \textit{quantization} to the process of transforming the \gls*{ECG}, which is a continuous signal, into a discrete one \cite{Calleja2015,Rostami2013}. There are three main quantization algorithms \cite{Calleja2015}: scalar, uniform, and dynamic. 
\begin{compactenum}
    \item[Scalar Quantization Algorithm.] Directly transforms the extracted \glspl*{IPI} into integers representation, typically represented in an 8-bit unsigned integer. 
    
    \item[Uniform Quantization Algorithm.] It implements the same functionality as the scalar one with the only difference that the output is mapped into an integer of a given precision.
    
    \item[Dynamic Quantization Algorithm.] This algorithm assumes that the signal has noise which can be modelled using a normal distribution with mean $\mu=0$ and a standard deviation $\sigma$ usually in the range $[0,1]$. Finally, in order to generate values of a fixed length, these values should be multiplied by a factor, \eg\ multiplying the values by 256 ($2^8$), the final output will have 8-bit unsigned integers. 
\end{compactenum}

Given the similarities of both the scalar and the uniform quantization algorithms, we simplify the quantization algorithms in two main families: uniform and dynamic.

\vspace{5mm}

\begin{figure}
    \centering
    \includegraphics[width=\linewidth]{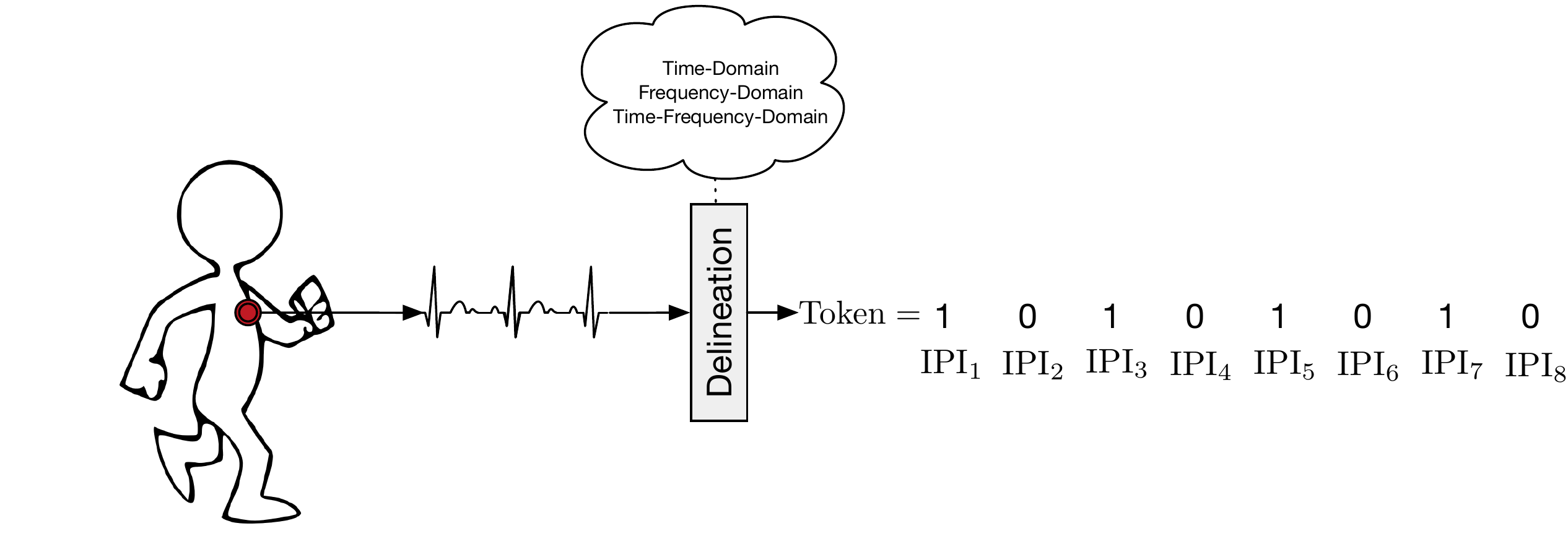}
    \caption{8-bits token generation example.}
    \label{fig:8_bits_generation}
\end{figure}

In \Cref{fig:8_bits_generation}, we include an example of an 8-bits token generation based on the heart signal. We can see how the sensor gathers the heart signal and, by applying a delineation algorithm, the time difference between R-peaks (\glspl*{IPI}) is transformed into a discrete signal. Later, we can generate tokens composed of 8 bits by, for instance concatenating the bits of the different \glspl*{IPI}. 

In the following, we explain in more detail each one of the delineation algorithms used to transform the continuous signal to a discrete one. Most of these algorithms are coded in Python and freely available \cite{luis_2019_3588108}. For more information, we refer to a survey published by Oweis and Al-Taaba \cite{Oweis2014} in 2014 where authors review and compare different methods about how to detect fiducial points in the \gls*{ECG} signal.

\subsection{Delineation of ECG: Time Domain}\label{subsec:time_domain}


We can classify time-domain delineation algorithms based on the filters they use. Therefore, before going forward, let us explain the differences between the main two families of filters: \gls*{FIR} and \gls*{IIR}. The former are distinctive in that their impulse response lasts for a finite time duration whereas the latter has an impulse response that is infinite in duration. 

\gls*{FIR} filters are usually easier to implement than \gls*{IIR} filters. Some examples are raised cosine, root raised cosine, cosine, sine, and matched filters. On the contrary, \gls*{IIR} filters are known by their efficiency in implementation, specifically in terms of passband, stopband, ripple, and roll-off. Some examples are Bessel, Chebyshev II, Butterworth, delay, Chebyshev I, Elliptic, Gaussian, Hourglass, Legendre, and matched.


Even though the fiducial points extraction in the time domain of the heart signal is an ongoing research topic \cite{Lee2018}, one of the most common procedures to extract information from the \gls{ECG} is the Pan–Tompkins's QRS detection algorithm \cite{pan1985real}. With it, not only the R-peaks can be extracted but also other peaks like the Q, R, and S. The accuracy of this algorithm was tested against the mitdb database \cite{mitdb} of Physionet repository, achieving 99.3\% of a correctly detected QRS complex.

Despite Pan-Tompkins being the most popular algorithm used to detect R-peaks in the literature, in the following, we briefly review alternative algorithms that might also be used to extract fiducial points in the time domain. 

\paragraph{\textbf{Pan–Tompkins}}
In summary, the Pan–Tompkins algorithm that consists of five steps, 
\begin{inparaenum}[1)]
    \item band and pass filter (low and high filters);
    \item derivatives;
    \item squaring function, and;
    \item moving-window integration. 
\end{inparaenum}
The low pass filter reduces the influence of muscle noise, 60Hz interference, baseline wander, and T-wave interference. The high pass filter subtracts the output from the first-order low-pass filter with an all-pass filter output. These filters are usually implemented as an \gls*{IIR} filter (Butterworth filter) 5-15Hz. The derivative differentiates the signal to produce the slope of the QRS complex and after that, the signal is squared point-by-point in squaring function to avoid T-peaks being considered as R-peaks (all the points will be positive at this point). Finally, the moving window (150ms) integration will give the slope of the R waveform.

\paragraph{\textbf{Hamilton}} Patrick Hamilton proposed in 2002 a delineation algorithm based on the Pan–Tompkins one \cite{Hamilton2002}. The main difference that he introduced is  using a band-pass of 8-16Hz instead of squaring it. Also, the algorithm uses a different moving window of 80ms instead of the 150ms that the original Pan–Tompkins used.

\paragraph{\textbf{Christov}} Igor Christov proposed a method to detect the QRS complex wave as well as the R-peaks based on 3 moving windows to manipulate the signal and a combination of 3 adaptive thresholds to detect the QRS complex.

\paragraph{\textbf{Engelse and Zeelenberg with modifications by Lourenco et al}} Lourenco \etal\ \cite{Lourencco2012} expanded the algorithm previously proposed by Engelse \etal\ \cite{Engelse1979} to make it deployable in real-time scenarios. To do so, they replaced the originally fixed threshold with an adaptive one. This algorithm, after filtering the noise (band-stop filter) it uses the delineation algorithm proposed by Christov \cite{Christov2004} to model the adaptive threshold. Finally, they assume that within 160ms of a peak being detected there are at least 10ms of consecutive points, being the R-peak the maximum value in this time window.

\paragraph{\textbf{Two Moving Average}}
Also using Pan-Tompinks as a basis, Elgendi \etal\ \cite{Elgendi2010} proposed in 2010 a Butterworth IIR filter with a band-pass of 8-20Hz, \ie\ having a frequency response as flat as possible in the band-pass and two moving averages. The first one with a window of 120ms tries to match the duration of a QRS complex whereas a second wider window of 600ms tries to match the approximate duration of a heartbeat.

\paragraph{\textbf{Matched Filter}}
This method uses the same threshold method introduced by Pan-Tompkins algorithm plus a filter to match an \gls*{ECG} to a QRS template. In more detail, the heart signal is first filtered using a bandpass filter (0.1-48Hz) to remove both DC and 50Hz power-line noise. After that, the template is loaded and the time-reversed as coefficients so the signal is then filtered. Finally, and similar to the Pan–Tompkins algorithm, the output of the filter is squared and the R-peaks retrieved.

\subsection{Delineation of ECG: Frequency Domain}\label{subsec:frequency_domain}

The \glsentryfull{FT} decomposes the signal in infinite sinusoidal functions. Therefore, all the time fiducial points like the QRS complex wave are spread over the frequency domain. In other words, when we apply \gls*{FT} to \glspl*{ECG}, we decompose the heart signal into the frequencies that compose it by computing an inner product of basis functions with the signal ($e^{-i\omega t}$). 

More formally, the \gls*{FT} of a heart signal $x(t)$ is given as:

\begin{equation*}
    FT(x(t))(\omega)=\int_{-\infty}^{+\infty}x(t)e^{-i\omega t}dt, \text{ where } \omega \text{ is the frequency}.
\end{equation*}

The \gls*{FT} assumes the signal to be stationary. However, this is not the case of \gls*{ECG} signals. To by-pass this restriction, a short enough time window of the signal is usually analyzed such that we can assume it is stationarity. This is done by using the \gls*{STFT} that includes the time dimension $m$ ($t-t_0$) to the base function by integrating a window of the complex exponential. Formally:

\begin{equation*}
    \begin{array}{l}
        STFT(x(t))(\omega, m)=\int_{-\infty}^{+\infty}x(t)\omega(m)e^{-i\omega t}dt, \text{ where } \omega \\ \text{ is the frequency, and } m \text{ the time window}.
    \end{array}
\end{equation*}

For the window functions, there can be used different types of functions like Kaiser, rectangular, Bartlett, Hanning, Hamming, and Blackman among others to reduce the effects of spectral leakage at the first side lobe \cite{Harris1978}.

As an example, Hu \etal\ \cite{Hu2013} use a feature extraction method based on \gls*{FFT} but also they are concerned about the computational demanding operation that \gls*{FFT} has in wearables and medical devices and use an \gls*{IPI}-based algorithm to extract common tokens in different devices. It is also important to remark that, despite some authors use \gls*{FFT} to extract features from the \gls*{ECG}, they also use a quantization method later on to transform the acquired \glspl*{IPI} into tokens from a given length \cite{Bao2013,Venkatasubramanian2010,Venkatasubramanian2010b,Xu2011}.

\subsection{Delineation of ECG: Time-Frequency Domain}

The \glsentryfull{WT} provides a description of the signal. By applying a linear transform, the \gls*{WT} decomposes a signal into different components at different scales, \ie\ different frequencies. More formally, a wavelet is essentially used to refer to a family of basis functions of the Hilbert space $L^2(\mathbb{R}^n)$, generated from a finite set of normalized functions $\psi_i$ where $i$ is chosen from a finite set $I$, and two operations: scaling (a), and translation (b). More concretely, the \gls*{WT} of a signal $x(t)$ is:

\begin{equation*}
    W_ax(b)=\frac{1}{\sqrt{a}}\int_{-\infty}^{+\infty}x(t)\psi \left( \frac{t-b}{a}\right) dt,\: a>0
\end{equation*}

We can distinguish between two \glspl*{WT}: \gls*{CWT} and \gls*{DWT}. \gls*{CWT} is usually generated by letting both the translation and scale operations vary continuously. \gls*{DWT} uses a pair of filters to successively isolate both low and high pass components of a signal \cite{Torrence1998}. Hence, due to the non-stationary nature of the \gls*{ECG}, the \glspl*{DWT} are especially suitable for such a signal and the continuous repetition of its patterns/waveforms, \eg\ QRS complex or P and T waves, at different frequencies \cite{Martinez2004,Banerjee2012}.

Mart\'inez \etal\ \cite{Martinez2004} provide a detailed explanation about how, by taking as a prototype wavelet ($\psi_i$) a smoothing function, discretizing either or both parameters $a$ or $b$ and taking a dyadic grid on the time-scale plane such that $a=2^k$ and $b=2^kl$, where $k$ controls the dilation or translation and $l$ the position of the wavelet function, then the transform is then called \textit{dyadic wavelet transform}, with basis functions:

\begin{equation*}
    \psi_{k,l}(t)=2^{-k/2}\psi(2^{-k}t-l); \: \text{where}\: k,l\in \mathbb{Z}^+
\end{equation*}

Roughly speaking, the \gls*{DWT} analyzes the signal at different resolutions through the decomposition of the signal into several successive frequency bands. To do so, two set functions are typically used: $\psi_{k,l}(t)=2^{-k/2}\psi(2^{-k}t-l)$ and $\phi_{k,l}(t)=2^{-k/2}\psi(2^{-k}t-l)$, normally linked with the low/high pass filters respectively \cite{Daubechies1990}. 

\gls*{DWT} has several families like Coiflet, Daubechies, Biorthogonal, Haar, or Symlet among others \cite{Meurant2012}. Even though it is proven that Daubechie order 6 performs better than other families \cite{Mahmoodabadi2005,Moosavi2017b}, we find authors who use the Daubechie order 4 wavelet (DB4) family \cite{Moosavi2017,Camara2018}, or; Daubechie order 6 and 9 wavelets (DB6, DB9) \cite{Moosavi2017b}.

\begin{figure}
\centering
\includegraphics[width=.8\columnwidth]{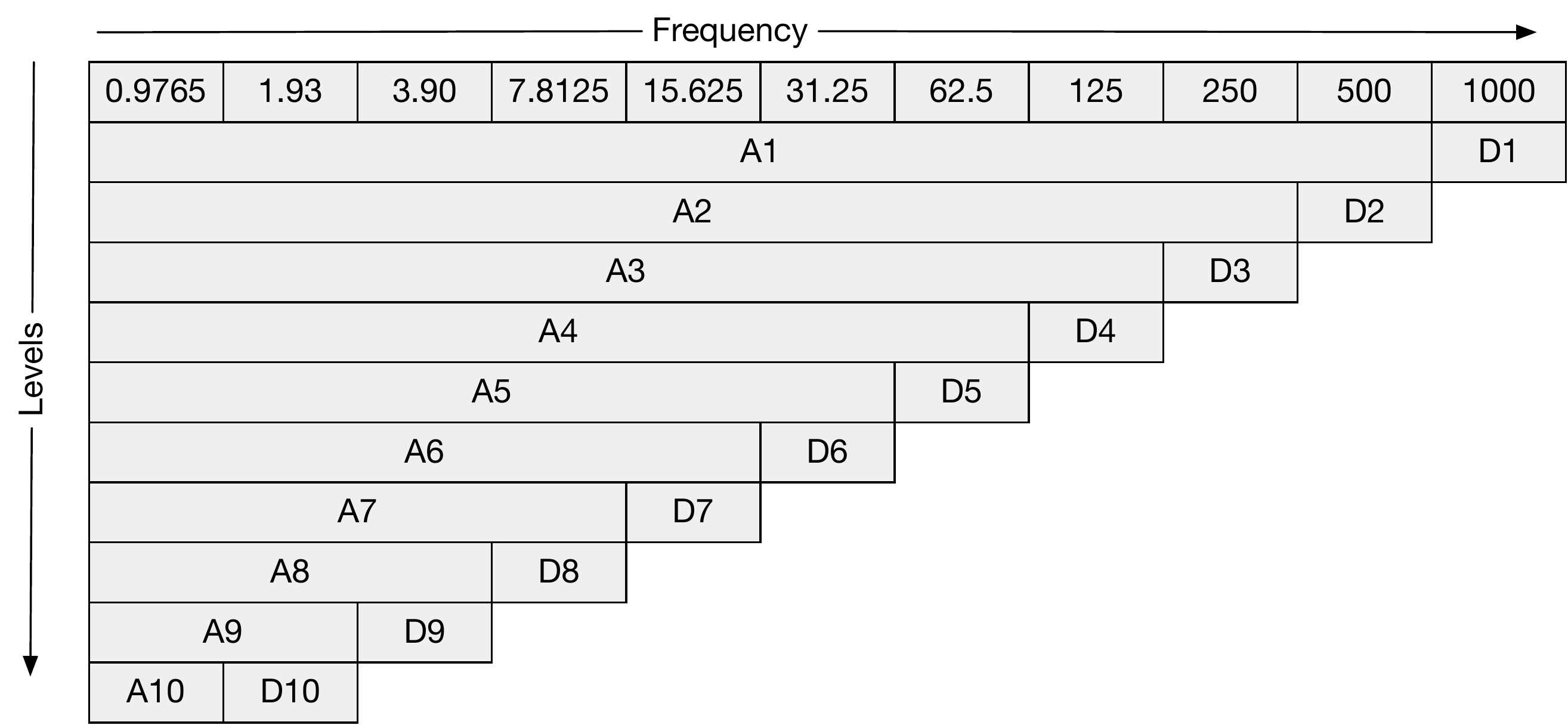}
\caption{DWT decomposition of a signal sampled at 1kHz \cite{Banerjee2012}.}
\label{figure:dwt_decomposition}
\end{figure}

Let us explain the \gls*{DWT} concept with an example applied to \gls*{ECG} signals.  \Cref{figure:dwt_decomposition} shows a decomposition using wavelets of a signal $x(t)$ sampled at 1kHz (like iafdb \cite{iafdb} or ptbdb \cite{ptbdb} databases). This decomposition is repeated up to 10 times to increase the frequency resolution as well as the approximation coefficients decomposed with both high ($\psi_{k,l}$) and low ($\phi_{k,l}$) pass filters. Note that \Cref{figure:dwt_decomposition} can also be represented as a binary tree, known as filter bank \cite{Flandrin2004}, where nodes represent sub-spaces with different time-frequencies of the signal. To know more about wavelets, we urge readers to a well-known book \cite{Meurant2012} and some Matlab implementations\footnote{\url{https://uk.mathworks.com/help/wavelet/index.html}}.

\subsection{Summary of Surveyed Papers}

\Cref{table:bits_authors} shows a summary of the surveyed publications. We added the year when they were published; the extraction algorithm used for \glspl*{IPI} generation; the number of extracted information in bits, and; the quantization algorithm (when stated). 

Concerning the delineation algorithm: 
\begin{inparaenum}[1)]
    \item Zaghouani \etal\ \cite{Zaghouani2015} use a \gls*{DCTAC} \cite{Plataniotis2006} of heartbeat records to process the signal and extract features as well as a \gls*{LPC} to interchange the derived token from the \glspl*{IPI} among sensors in the \gls*{WBAN}. An in-depth analysis of \gls*{LPC} was reviewed by Makhoul \cite{Makhoul1975};
    \item Chen \etal\ \cite{Chen2012} modeled the \gls*{ECG} signal as a dynamic system and applied a Lyapunov exponent’s spectrum to extract the features of the signal to be used as a secret key in a cryptographic protocol.
    \item González-Manzano \etal\ \cite{Manzano2017} use a Walsh–Hadamard Transform---which is a special subclass of \glspl*{FT}, to extract information from the \glspl*{ECG} to be used afterward to encrypt data.
    \item Bai \etal\ \cite{Bai2019} propose to use multiple features from the \gls*{ECG} to generate a secret key with the help of a \gls*{LFSR}.
\end{inparaenum}

Regarding the number of extracted bits, the majority of the surveyed papers extract 4 bits of information per \gls*{IPI}, however, this tendency is changing since 2016 in favor of extracting more information  (see \Cref{table:bits_authors}). Only a few authors \cite{Altop2017,Kim2018,Koya2018,Reshan2019,Lin2019} still extracted less than 4 bits whereas the majority of them \cite{Manzano2017,Moosavi2017,Seepers2017,Pirbhulal2018,Camara2018,Karthikeyan2018} extracted more information from the heart signal (due to the use of \gls*{WT} and \gls*{DWT} delineation algorithms). 
In more detail, we found in the literature a wide disparity in this matter. There are authors who:
\begin{inparaenum}[i)]
    \item do not mention the number of taken bits (or it is difficult to infer from the explanations) \cite{Bao2004,Bao2005,yao2011biometric,Zaghouani2015,Ingale2020,Mahendran2020,Belkhouja2019,Zheng2019,Sammoud2018};
    \item extract 2 bits \cite{Hong2011,Poon2006,Lin2019};
    \item vary the number of bits \cite{Bao2013,Zhang2012};
    \item take 4 bits \cite{Altop2017,Kim2018,Koya2018,Rostami2013,Seepers2015,Xu2011,Reshan2019};
    \item use between 5 and 8 bits \cite{Vasyltsov2016,Zheng2015,Bao2008,Seepers2017};
    \item take 16 bits \cite{Pirbhulal2018,Zheng2016}, or;
    \item use more than 16 bits \cite{Camara2018,Karthikeyan2018,Bai2019}.
\end{inparaenum}

\begin{table}[t]
\caption{Summary of the surveyed papers with the year of publication (column 2); the extraction methodology (column 3); the number of bits extracted from IPIs (column 4), and; the quantization algorithm (column 5). ``---'' means that either no info is provided or we could not directly derive it from the paper. }
\label{table:bits_authors}
\centering
\resizebox{\columnwidth}{!}{%
\begin{tabular}{l r l l l}
\toprule
   & \textbf{Year} & \textbf{Extraction} & \textbf{Bits per \glspl*{IPI}} & \textbf{Quantization}\\ 
\midrule
  Ingale \etal\ \cite{Ingale2020} & 2020 & \gls*{WT} & --- & --- \\
  Mahendran \etal\ \cite{Mahendran2020} & 2020 & \gls*{DWT} & --- & ---\\
  Bai \etal\ \cite{Bai2019} & 2019 & --- & 16 bits & ---\\
  Belkhouja \etal\ \cite{Belkhouja2019} & 2019 & --- & --- & ---\\
  Lin \etal\ \cite{Lin2019} & 2019 & \glspl*{IPI} & 3 bits & Dynamic\\
  Reshan \etal\ \cite{Reshan2019} & 2019 & \glspl*{IPI} & 4 bit & Dynamic \\
  Zheng \etal\ \cite{Zheng2019} & 2019 & \glspl*{IPI} & --- & Uniform \\
  Camara \etal\ \cite{Camara2018}& 2018 & \gls*{DWT} & 184 bits & ---\\
  Karthikeyan \etal\ \cite{Karthikeyan2018}& 2018 & \gls*{WT} & 64 bits & ---\\
  Pirbhulal \etal\ \cite{Pirbhulal2018}& 2018 & \glspl*{IPI} & 16 bits & Uniform\\ 
  Kim \etal\ \cite{Kim2018}& 2018 & \glspl*{IPI} & 4 bits & Uniform\\ 
  Koya \etal\ \cite{Koya2018} & 2018 & \glspl*{IPI} & 4 bits & Uniform\\
  Sammoud \etal\ \cite{Sammoud2018} & 2018 & --- & --- & ---\\
  Wu \etal\ \cite{Wu2018} & 2018 & \glspl*{IPI} & 8 to 12 bits & Uniform \\
  González-Manzano \etal\ \cite{Manzano2017} & 2017 & \glspl*{FT} & up to 300 bits & --- \\
  Moosavi \etal\ \cite{Moosavi2017b}& 2017 & \glspl*{DWT} + \glspl*{IPI} & $\approx$16 bits & Dynamic\\ 
  Moosavi \etal\ \cite{Moosavi2017}& 2017 & \glspl*{DWT} + \glspl*{IPI} & 8 bits & ---\\ 
  Seepers \etal\ \cite{Seepers2017}& 2017 & \glspl*{IPI} & 8 bits & Uniform\\
  Sun \etal\ \cite{Sun2017} & 2017 & \glspl*{IPI} & 2 bits in 50Hz & Dynamic \\
  Zaghouani \etal\ \cite{Zaghouani2017,Zaghouani2017b} & 2017 & \gls*{DCTAC} & --- & ---\\
  Altop \etal\ \cite{Altop2017} & 2017 & \glspl*{IPI} & 4 bits & Uniform\\
  Zheng \etal\ \cite{Zheng2016} & 2016 & \glspl*{DWT} & 16 bits & Dynamic \\ 
  Peter \etal\ \cite{Peter2016} & 2016 & \glspl*{IPI} & --- & Uniform\\
  Vasyltsov \etal\ \cite{Vasyltsov2016} & 2016 & \glspl*{IPI} & 5 bits & Uniform\\
  Pirbhulal \etal\ \cite{Pirbhulal2015} & 2015 & \glspl*{IPI} & 16 bits & Dynamic \\
  Zheng \etal\ \cite{Zheng2015} & 2015 & \glspl*{IPI} & 7 bits & Uniform\\ 
  Seepers \etal\ \cite{Seepers2015}& 2015 & \glspl*{IPI} & 4 bits & Uniform\\
  Zaghouani \etal\ \cite{Zaghouani2015} & 2015 & \gls*{DCTAC} & --- & ---\\
  Seepers \etal\ \cite{Seepers2014}& 2014 & \glspl*{IPI} & 4bits & Uniform\\
  Bao \etal\ \cite{Bao2013} & 2013 & \glspl*{IPI} & 3 to 4 bits & Uniform\\
  Rostami \etal\ \cite{Rostami2013}& 2013 & \glspl*{IPI} & 4 bits & Dynamic\\ 
  Hu \etal\ \cite{Hu2013} & 2013 & \glspl*{FFT} + \glspl*{IPI} & 4 bits & Uniform\\ 
  Zhang \etal\ \cite{Zhang2012} & 2012 & \glspl*{IPI} & 2 to 11 bits & Uniform\\ 
  Chen \etal\ \cite{Chen2012} & 2012 & Lyapunov exponents & --- & ---\\ 
  Xu \etal\ \cite{Xu2011} & 2011 & \glspl*{WT} + \glspl*{IPI} & 4 bits & Dynamic\\ 
  Hong \etal\ \cite{Hong2011} & 2011 & \glspl*{IPI} & 2 bits &Uniform\\
  Yao \etal\ \cite{yao2011biometric} & 2011 & \glspl*{WT} & --- & ---\\
  Venkatasubramanian \etal\ \cite{Venkatasubramanian2010,Venkatasubramanian2010b} & 2010 & \gls*{FFT} & 1 to 15 bits & ---\\
  Zhang \etal\ \cite{Zhang2010} & 2010 & \glspl*{IPI} & --- & Uniform\\ 
  Bao \etal\ \cite{Bao2008} & 2008 & \glspl*{IPI} & 4 bits & Uniform\\
  Poon \etal\ \cite{Poon2006} & 2006 & \glspl*{IPI} & 2 to 7 bits & Uniform\\
  Bao \etal\ \cite{Bao2005} & 2005 & \glspl*{IPI} & --- &Uniform\\
  Bao \etal\ \cite{Bao2004} & 2004 & \glspl*{IPI} & --- & Uniform\\
\bottomrule
\end{tabular}%
}
\end{table}
\section{Randomness of the Heart Signal}\label{sec:tests}

Randomness is a needed property in cryptography because it provides a way to create information that an adversary cannot learn or predict \cite{Gennaro2006}. Several authors \cite{Altop2017,Seepers2015,Seepers2017,Pirbhulal2019,Bao2004,Karthikeyan2018,Koya2018,yao2011biometric,Moosavi2017,Seepers2015} demonstrated that the \gls*{ECG} signal can be used as a source of randomness due to its high entropy and thus, \textit{nonces} or random numbers can be derived from it. However, according to Rushanan \etal\ \cite{Rushanan2014}, running Shannon's entropy test or any variant of it (\eg\ Collision or Rényi \cite{Renyi1961}), is not enough to claim that the \gls*{ECG} can be a good source of entropy.  

To test the random property of a dataset, we can find test suites like AIS31 \cite{Killmann2011}, \glsentryfull{NISTSTS}, and ENT. Indeed, many of the papers we surveyed run some (parts) of these suites. In the following, we explain the tests that these suites are composed of as well as the constraints they have, if any.

\subsection{Suites to Test Random Tokens}

\paragraph{AIS31} Is a suite defined by the German Federal Office for Information Security (BSI)\footnote{\url{https://www.bsi.bund.de/EN/Home/home_node.html}} \cite{Killmann2011}. It is composed of eight tests: disjointness, monobit, poker, run, longrun, autocorrelation, uniform distribution, comparative test for multinomial, and entropy. What makes this suite different from the others is that all the tests are designed to be performed on the randomness source instead of running over the generated tokens \cite{Park2016,Guilley2018}. A more detailed description of each test can be seen in \Cref{table:AIS31}.

\paragraph{\gls*{NISTSTS}} The \gls*{NISTSTS} \cite{Bassham:2010:SRS:2206233} is a set of fifteen statistical tests to evaluate random and pseudo-random number generators used in cryptographic applications. 
They are often used as a first step in spotting low-quality generators, but successfully passing all tests does not guarantee that the generator is strong enough. This suite essentially inherits and extends most of the Diehard suite tests \cite{Marsaglia2008} but without the size constraint that it has (more than 80 million of bits). We include a description of each test in \Cref{table:NIST}.

\paragraph{ENT} This suite provides a comprehensive analysis of randomness of a sequence of bits. It is composed of five tests: entropy, Chi-square ($\chi^2$), arithmetic mean ($\bar{x}$), Monte Carlo value for $\pi$, and serial correlation coefficient. A more detailed description of each test can be seen in \Cref{table:ENT}.

\begin{algorithm}
	\SetKw{PanTomkins}{panTomkins}
	\SetKw{calculateIPI}{calculateIPI}
	\SetKw{quantization}{quantization}
	\SetKw{Grey}{Grey}
	\SetKw{len}{len}
	\SetKw{break}{break}
	\KwIn{ECG, frequency, tokenLength}
	\KwOut{token}
	\BlankLine
	 peaks = \PanTomkins(ECG, frequency);\\
	 IPIs = \calculateIPI(peaks);\\
	 IPIs = \quantization(IPIs);\\
	 token = []\\
	 \ForEach{counter in IPIs.len()}{
	    IPIs[counter] = \Grey{(IPIs[counter])};\\
	    \If{token.len()$<$tokenLength}{
	        token += IPIs[counter][5:8];\\
	 }\lElse{return token[0:tokenLength]}
	 }
	\caption{Token generation algorithm taking the 4 LSBs of the IPIs.}\label{alg:ipiExtraction}
\end{algorithm}


\textbf{Example 1: IPI-based Token Generation.} We included in  \Cref{alg:ipiExtraction} a pseudo-code of a cryptographic token based on \glspl*{IPI}. The algorithm needs two input values: the signal and the frequency of the signal. 
First, the algorithm runs the Pan-Tomkins's QRS detection algorithm \cite{pan1985real} to extract the R-peaks from the \gls*{ECG} signal. Second, 
we compute the time difference between each pair of two consecutive R-peaks, \ie\  \glspl*{IPI}. Third, for each \gls*{IPI} we apply any of the aforementioned quantization algorithms. Note that this process consists of generating discrete values from the \gls*{ECG} continuous signal.

Once we quantized the \glspl*{IPI}, we:
\begin{inparaenum}[1)]
    \item apply a Grey code to increase the error margin of the physiological parameters (line 6 of \Cref{alg:ipiExtraction}), and;
    \item extract the 4 \gls*{LSB} from each coded \gls*{IPI} value (line 8 of \Cref{alg:ipiExtraction}).
\end{inparaenum} For the sake of simplicity, we included here the same coded include in the original proposal, however, the amount of bits derived from the \glspl*{IPI} can easily be modified.


\subsection{Heart as a Source of Entropy: NIST SP 800-90B Recommendation}

Random number generators need some noise source to produce outputs with a certain level of unpredictability. \gls*{NIST} proposed a set of recommendations in a special publication named SP 800-90B \cite{NIST80090} to evaluate the entropy sources. These entropy sources together with random-bit generator mechanisms (specified in the special publication NIST SP 800-90A) can be used to create random-bit generators (following the NIST SP 800-90C recommendations).

To quantify the quality of a source of entropy, \gls*{NIST} introduces the concept of \textit{min-entropy} as a battery of 10 tests or estimators. Each one of these estimators outputs an entropy value independent from the other estimators. The final min-entropy value is the minimum value of each one of the estimators. The min-entropy differs from the one used in information theory---which is a specific case of R\'{e}nyi's entropy where the uncertainty is measured in terms of a random variable’s vulnerability to being guessed in one try by an adversary \cite{Renyi1961}.

NIST released a public implementation of the 10 estimators \cite{NIST80090Btests}: 
\begin{inparaenum}[1)]
    \item the most common value estimate;
    \item the collision estimate;
    \item the Markov estimate;
    \item the compression estimate;
    \item the multiMCW prediction estimate;
    \item the lag prediction estimate;
    \item the multiMMC prediction estimate;
    \item the LZ78Y prediction estimate;
    \item the t-Tuple estimate, and;
    \item the LRS estimate.
\end{inparaenum} See \Cref{table:NIST80090} for a more detailed description of each one of the estimators (\ref{sec:appendix}).

\subsection{Summary of Surveyed Papers}

We include in \Cref{table:random_tests_authors} a summary of the main random tests that the surveyed papers use in their proposals to validate their results. The first column corresponds to the method or suite; the second one stands for more detailed information (if any) about the test or the suite. For instance, we split the \gls*{NIST} into the number of different tests that the proposals use: 4, 5, 6, 8, 9, 10, and 15 tests. Finally, in the third column, we include those papers that match the previous information. Note that the same paper might use different methods, \eg\ Bao \etal\ \cite{Bao2008} run 5 tests of the \gls*{NIST} suite as well as the entropy test.

Surprisingly, most of the papers that use any of the suites to measure the randomness of the tokens derived from the heart signal do not execute all of the tests that the suites are composed of but Camara \etal\ \cite{Camara2018}. Unfortunately, we could not find a pattern among the surveyed papers about which suites, tests, or sub-tests the authors run to evaluate the randomness tokens. 

Remarkably, there is a large number of works (16) that do not evaluate how good the derived tokens are. In addition to that, we did only find two papers where authors evaluated whether the heart can be considered a good source of entropy \cite{Ortiz2020,Chizari2019}. The main reason might be due to the high constraints in terms of size of the dataset that the \gls*{NIST} set: \say{a sequential dataset of at least 1,000,000 consecutive sample values obtained directly from the noise source}. Because these constraints might not be easily fulfilled, \gls*{NIST} also allows the concatenation of chunks of 1,000 samples to generate a final dataset of 1,000,000 samples. 


\paragraph{\textbf{Questioning the Randomness of Heart-based Tokens}} Cherukuri \etal\ in 2003 \cite{Cherukuri2003} were the first who questioned the use of the \gls*{ECG} as a source of entropy. Later, in 2018 Ortiz \etal\ \cite{Ortiz2018} run both \gls*{NIST} and ENT suites against 19 databases from Physionet and concluded that the \glspl{IPI} extracted from \gls*{ECG} using the dynamic algorithm proposed by Rostami \etal\ \cite{Rostami2013} cannot be considered to be random. A similar conclusion was reached by Tuncer \etal\ \cite{Tuncer2018} later on that year. 

Note that these works demonstrate that the delineation algorithms used to generate tokens from the heart signal are not optimal enough and the tokens are not totally random. In other words, this does not mean that the heart signal is not random.

\begin{table}[ht]
\caption{Random tests applied, if any, to the tokens derived from the heart signal. }
\label{table:random_tests_authors}
\centering
\resizebox{\linewidth}{!}{%
\begin{tabular}{l l p{4cm}}
\toprule
  Random Test(s) & Subtest(s) & Reviewed Papers\\ 
\midrule
  \multirow{7}{*}{\gls*{NIST} (15 tests)} & 4 tests & \cite{Belkhouja2019} \\
  & 5 tests & \cite{Bao2008,Moosavi2017,Moosavi2017b,Zhang2010,Zhang2012} \\
  & 6 tests & \cite{Hong2011} \\
  & 8 tests & \cite{Kim2018,Lin2019,Rostami2013,Zheng2015,Zheng2019} \\
  & 9 tests & \cite{Pirbhulal2018,Wu2018,Xu2011} \\
  & 10 tests& \cite{Zheng2016} \\
  & 15 tests& \cite{Camara2018} \\ 
  \\
  \multirow{4}{*}{Entropy} & Min-Entropy & \cite{Manzano2017,Vasyltsov2016}\\ 
  & Shannon & \cite{Altop2017,Moosavi2017,Moosavi2017b,Pirbhulal2018,Seepers2014,Seepers2015,Seepers2017,Vasyltsov2016,Zhang2012}\\
  & Rényi & \cite{Vasyltsov2016}\\
  & Entropy & \cite{Bao2004,Bao2008,Karthikeyan2018,Koya2018,Lin2019,Moosavi2017,Poon2006,Wu2018,Zheng2015,Zheng2016,Zheng2019}\\ 
  \\
  AIS.31 (9 tests) & 5 tests & \cite{Kim2018}\\ 
  DIEHARD (15 tests) & 15 tests & \cite{Camara2018}\\ 
  \\
  \multirow{2}{*}{ENT (6 tests)} & 3 tests & \cite{Seepers2015}\\ 
  & 6 tests & \cite{Camara2018}\\ 
  \\
  \multicolumn{2}{l}{Chi-Square} & \cite{Bao2004,Moosavi2017b,Poon2006}\\
  \multicolumn{2}{l}{No tests} & \cite{Bao2005,Bao2013,Bai2019,Chen2012,Hu2013,Mahendran2020,Peter2016,Pirbhulal2015,Reshan2019,Sun2017,Venkatasubramanian2010,Venkatasubramanian2010b,yao2011biometric,Zaghouani2015,Zaghouani2017,Zaghouani2017b}\\
\bottomrule
\end{tabular}%
}
\end{table}
\section{Challenges and Conclusions}\label{sec:conclusions}

\textbf{Challenges and Directions.}
We identified some open questions as well as future research directions. In the following, we detail a few of them:
\begin{itemize}
    \item When authors use their own sensors, either created modified by them, it is, in general, difficult to evaluate whether a cryptographic protocol is good or bad in comparison to others due to the proposal itself or because of the sensors they use. We then argue that cryptographic protocols should use a common database to isolate and evaluate the proposed scheme.

    \item Databases of the Physionet repository are out-of-date, being majority gathered in the '90s using a low sample frequency--probably due to the sensors existing at that time. Some authors confirmed \cite{Ortiz2018} that databases sampled at high frequency like cebsdb produce better random numbers from \glspl*{IPI} when they are analyzed by some random suites. We advocate the need to record new databases sensed at high frequencies and to make these datasets public for the research community.
    
    \item \Cref{table:thew_description,table:physionet_description}, show that most databases count with twelve leads (standard diagnostic configuration) or three leads (common approach when wireless transmission is at play). Regarding the datasets, it would be advisable that authors clearly describe the placement of the electrodes---\eg\ well-known databases, such as nsrdb, do not count with that description. Besides, in many existing solutions built on public databases, the designers do not specify the lead used in their experiments which is critical to guarantee the reproducibility of the results.  
    
    \item There is no formal framework, guidelines, or set of (automatic) tests that authors might run to validate and compare their results with the existing proposals. 
    
    \item It is, in general, impossible to test the correctness of works based on proprietary data. We argue that all the data should be shared to allow other researchers to reproduce the same results and validate the proposed methodologies. If repositories are private, researchers have to conduct the same experiments repeatedly, and different results can be found for the same problem. Consequently, no one can be sure if the results extracted from the experiments are correct or better than others.
    
    \item There is a lack of consensus about how long the heart signal should be sensed to derive a cryptographic value. Even when the protocol needs 30 seconds to generate a valid token, depending on the scenario, this might be infeasible, \eg\ a patient suffering from a heart attack. Better and faster protocols need to be proposed to cover most of the scenarios where \glspl*{IMD} work.
    
    \item Sometimes, a nonce does not need to be ideally generated from the cryptographic point of view, \ie\ it does not need to pass all the security and random tests. Depending on the application, different levels of security should be considered. Therefore it might be useful to define what the minimum conditions are to be used in different scenarios. 
    
    \item We can see how researchers are using more bits than the traditional four \gls*{LSB} by using delineation algorithms based on the frequency domain like the \gls*{WT}. However, we could not check how these algorithms affect the consumption (power and computation) of the devices. This is, from our point of view, an important research direction since it will 1) define the minimum requirements of the sensors concerning the methodology used, and; 2) provide a lower bound of the lifetime of the batteries, being this crucial in the case of the \glspl*{IMD}.
    
    \item Related to the previous point, authors use many different environments to deploy their proposals. While most of the surveyed literature use theoretical or simulated environments, there are only a few works that deployed their solutions in real devices like a Raspberry Pi \cite{Karthikeyan2018,Ortiz2019} and a manufacturer's pacemaker \cite{Khoa2013}. Thus, we argue for more realistic scenarios to test the experiments, being this more important when the target are \glspl*{IMD}.
\end{itemize}

\textbf{Conclusions.}
This paper reviewed and classified most of the literature we found over the last 18 years on the heart signal and cryptographic protocols. We have explained the most important algorithms to transform the continuous heart signal into a discrete signal used afterwards in security protocols. We classified the existing proposals according to three primary parameters: 
\begin{inparaenum}[1)]
    \item the dataset they use for testing their results;   
    \item the delineation algorithms they use to extract the fiducial points, and;
    \item the cryptographic tests they run (if any) to validate how random the extracted token is.
\end{inparaenum} We finally identified and presented some future research on this field. 

Summarizing, although there have been significant contributions in the last 18 years, we believe that this area has great potential and hope that this article will help to move it forward.  


\bibliographystyle{elsarticle-num}
\bibliography{Survey_citation}

\appendix
\section{}\label{sec:appendix}
\begin{table*}[!ht]
\caption{Description of the AIS31 random tests \cite{Kim2018,Guilley2018}.}
\label{table:AIS31}
\centering
\resizebox{\textwidth}{!}{%
\begin{tabular}{l p{9cm}}
\toprule
Test & Description \\
\midrule
T0 (Disjointness) & Subsequent members are pairwise different\\ 
T1 (Monobit) & Uniform test for bit sequence of length 20,000\\ 
T2 (Poker) & Goodness-of-fit test for the number of 4 bits block\\ 
T3 (Run) & Test for the number of run which has l-length\\ 
T4 (Longrun) & Check up the occurrence run of length $\ge$ 34\\ 
T5 (Autocorrelation) & Autocorrelation value of $\sum$ (i-th bit $\oplus$ I 5000-th bit) which is approximately 2500\\ 
T6 (Uniform distribution) & Uniform distribution test using ratio of 0's and 1's\\ 
T7 (Comparative test for multinomial) & Goodness-of-fit test for h blocks by comparison\\ 
T8 (Entropy) & Estimate entropy as minimum distance of blocks \\ 
\bottomrule
\end{tabular}%
}
\end{table*}

\begin{table*}[!ht]
\caption{Description of the tests included in NIST STS \cite{Bassham:2010:SRS:2206233}.}
\label{table:NIST}
\centering
\resizebox{\textwidth}{!}{%
\begin{tabular}{l p{9.8cm}}
\toprule
Test & Description \\
\midrule
T0 (Frequency Monobit) & Uniform test for bit sequence\\ 
T1 (Frequency Test within a Block) & For $M > 1$, checks if the frequency of ones in an $M$-bit block is approximately $M/2$.\\ 
T2 (Run) & Test for the number of run which has l-length\\ 
T3 (Longrun) & Similar to T2 with $M > 1$ blocks \\ 
T4 (Binary Matrix Rank) & $m \times n$ binary matrices over $GF(2)$ and checks whether the ranks are linearly dependent\\
T5 (Discrete Fourier Transform (Spectral)) & Calculate the DFT and  number of peaks $<5\%$ \\
T6 (Non-overlapping Template Matching) & Splits sequence into $M$ substrings of length $l$ and seeks for patterns.\\ 
T7 (Overlapping Template Matching) & Similar to T6 but using a sliding window that advances 1 bit at a time\\ 
T8 (Maurer's ``Universal Statistical'') & Detects if the sequence can be significantly compressed without loss of information \\ 
T9 (Linear Complexity) & Computes the linear complexity of the input sequence\\
T10 (Serial)& Calculate the frequency of all possible overlapping $M$-bit patterns\\
T11 (Approximate Entropy)& Generates the frequency of all possible overlapping $m$-bit pattern in a sequence\\
T12 (Cumulative Sums) & Zeroes are converted to negative ones and ones remain the same. the cumulative sum should be close to zero\\
T13 (Random Excursions) & The number of cycles having exactly $K$ visits in a cumulative sum random walk\\
T14 (Random Excursions Variant) & Similar to T13 but the goal is to detect deviations from the expected number of visits to various states in the random walk\\
\bottomrule
\end{tabular}%
}
\end{table*}

\begin{table*}[!ht]
\caption{Description of the ENT random tests \cite{entpseudorandom}.}
\label{table:ENT}
\centering
\resizebox{\textwidth}{!}{%
\begin{tabular}{l p{9cm}}
\toprule
Test & Description \\
\midrule
T0 (Entropy) & Amount of information of the sequence, expressed as a number of bits per character\\ 
T1 (Chi-square ($\chi^2$)) & Computes $\chi^2$ and indicates how frequently a truly random sequence would exceed the calculated value.\\ 
T2 (arithmetic mean ($\bar{x}$)) & Adding up all bytes in the sequence and dividing it by the sequence length (in bytes) \\ 
T3 (Monte Carlo value for $\pi$) & Estimates the value of $\pi$ through a standard Monte Carlo method using the input sequence\\ 
T4 (Serial correlation coefficient) & Capture correlations in the sequence by checking how much each byte in the stream depends upon the previous one.\\ 
\bottomrule
\end{tabular}%
}
\end{table*}

\begin{table*}[!ht]
\caption{Description of the NIST SP 800-90B estimators \cite{NIST80090}}
\label{table:NIST80090}
\centering
\resizebox{\textwidth}{!}{%
\begin{tabular}{l p{9cm}}
\toprule
Test & Description \\
\midrule
T0 (The Most Common Value Estimate) & It finds the proportion $p$ of the most common value in the dataset and then constructs a confidence interval for such a $p$.\\

T1 (The Collision Estimate) & Based on \cite{Hagerty2012}, this tests estimates the probability of the most-likely output value based on the collision times---number of repeated values. It outputs a low entropy estimate when the source has a significant bias toward a particular value while a higher entropy estimate is given for longer average time collisions.\\

T2 (The Markov Estimate) & It measures the dependencies between consecutive values---dependencies values. \\ 

T3 (The Compression Estimate) & It is based on the Maurer Universal Statistic \cite{Maurer1992} and it computes how much the dataset can be compressed. It first generates a dictionary of values and then computes the average number of samples required to produce an output based on that dictionary.\\

T4 (The MultiMCW Prediction Estimate) & It is composed of multiple  \gls*{MCW} sub-predictors, each of which aims to guess the next output, based on the last $n$ outputs. Each sub-predictor extracts the most often value in a window of $n$ outputs. This test was designed for cases where the most common value changes over time but remains relatively stationary over reasonable lengths of the dataset.\\ 

T5 (The Lag Prediction Estimate) & This test has several sub-predictors each of which predicts the next output based on a so-called \textit{lag}. It keeps a counter of the number of times that each sub-predictor was correct and uses the best sub-predictor to predict the next value.\\

T6 (The MultiMMC Prediction Estimate) & It is composed of multiple \gls*{MMC} sub-predictors. Instead of keeping the probability of a transition like in a Markov model, the predictors record the observed frequencies for transitions from one output to a subsequent output and makes a prediction based on the most frequently observed transition from the current output.\\

T7 (The LZ78Y Prediction Estimate) & The LZ78Y predictor is loosely based on LZ78 encoding with the Bernstein's Yabba scheme \cite{Salomon2004} for adding strings to the dictionary. It keeps a dictionary of strings and continues adding new strings to the dictionary until the dictionary has reached its maximum capacity.\\

T8 (The t-Tuple Estimate) & It checks the frequency of t-Tuples, \ie\ pairs, triples, etc., that appears in the dataset and produces an estimate of the entropy per sample based on the frequency of those t-tuples.\\

T9 (The \gls*{LRS} Estimate) & It estimates the collision entropy (sampling without replacement) of the dataset based on the number of repeated tuples within the input dataset.\\
\bottomrule
\end{tabular}%
}
\end{table*}

\end{document}